\RequirePackage{ifpdf}
\documentclass[hyper,letterpaper]{JHEP3}
\pdfoutput=1
\usepackage{amsmath,amssymb,amsfonts,bm}
\usepackage{cite}
\usepackage{graphicx, wrapfig}
\usepackage{multirow}
\usepackage{verbatim}
\usepackage{appendix}
\usepackage{amsmath,amsthm}

\usepackage{url}
\usepackage{float}
\usepackage{graphicx}

\newtheorem{theorem}{Theorem}[section]
\newtheorem{proposition}{Proposition}[section]
\newtheorem{lemma}[proposition]{Lemma}

\newtheorem{conjecture}[proposition]{Conjecture}

\def\printname#1{
        \if\draft y
                \smash{\makebox[0pt]{\hspace{-0.5in}
                        \raisebox{8pt}{\tt\tiny #1}}}
        \fi
}

\newlength{\standardunitlength}
\catcode`\@=11
\long\def\@makecaption#1#2{%
     \vskip 10pt

\setbox\@tempboxa\hbox{
       \small\sf{\bfcaptionfont #1. }\ignorespaces #2}%
     \ifdim \wd\@tempboxa >\captionwidth {%
         \rightskip=\@captionmargin\leftskip=\@captionmargin
         \unhbox\@tempboxa\par}%
       \else
         \hbox to\hsize{\hfil\box\@tempboxa\hfil}%
     \fi}
\font\bfcaptionfont=cmssbx10 scaled \magstephalf
\newdimen\@captionmargin\@captionmargin=2\parindent
\newdimen\captionwidth\captionwidth=\hsize
\catcode`\@=12

\newcommand{\bea}{\begin{eqnarray}}
\newcommand{\eea}{\end{eqnarray}}
\newcommand{\be}{\begin{equation}}
\newcommand{\ee}{\end{equation}}

\def\Tr{{\rm Tr \,}}

\def\disc{\mathrm{Disc}}

\def\BZ{\mathbb Z}
\def\BN{\mathbb N}
\def\BQ{\mathbb Q}

\renewcommand{\hat}{\widehat}

\title{Knots, BPS states, and algebraic curves}

\author{Stavros Garoufalidis$^{1}$, Piotr Kucharski$^2$ and Piotr Su{\l}kowski$^{2,3}$
\\
$^1$ School of Mathematics, Georgia Institute of Technology, Atlanta, GA 30332-0160, USA \\
$^2$ Faculty of Physics, University of Warsaw, ul. Pasteura 5, 02-093 Warsaw, Poland \\
$^3$ Walter Burke Institute for Theoretical Physics, California Institute of Technology, Pasadena, CA 91125, USA 
}

\abstract{We analyze relations between BPS degeneracies related to Labastida-Mari{\~n}o-Ooguri-Vafa (LMOV) invariants, and algebraic curves associated to knots.  
We introduce a new class of such curves that we call extremal A-polynomials, discuss their special properties, and determine exact and asymptotic formulas for the corresponding (extremal) BPS degeneracies. These formulas lead to nontrivial integrality statements in number theory, as well as to an improved integrality conjecture stronger than the known M-theory integrality predictions. Furthermore we determine the BPS degeneracies encoded in augmentation polynomials and show their consistency with known colored HOMFLY polynomials. Finally we consider refined BPS degeneracies for knots, determine them from the knowledge of super-A-polynomials, and verify their integrality. We illustrate our results with twist knots, torus knots, and various other knots with up to 10 crossings.
\\
\\
\\
\\
\\
\\
\\
\\
\\
\\
\\
\\
{\tt CALT-2015-021}}

\begin{document}

\tableofcontents


\newpage

\section{Introduction}
\label{sec.intro}

There are many profound relations between quantum field theory, string theory
and knot theory. This paper focuses on two aspects of the polynomial knot invariants -- in particular HOMFLY polynomials, superpolynomials, and their specializations -- colored by symmetric powers of the fundamental 
representation:
\begin{itemize}
\item
the special geometry of algebraic curves of a knot,
\item
the integrality of the BPS invariants of a knot.
\end{itemize}
By (affine) algebraic curves we mean curves obtained as classical limits  of recursions
of knot polynomials. Such recursions are known to exist (for colored Jones
polynomials \cite{GL,Hikami_AJ}, or colored HOMFLY and superpolynomials of twist or torus knots) and conjectured to exist for all colored HOMFLY or superpolynomials \cite{AVqdef,FGS,superA,Nawata}. For the colored Jones polynomial, the above algebraic curve agrees with the A-polynomial \cite{GL,Hikami_AJ,Apol}. For the colored 
HOMFLY polynomial, the above algebraic curve is conjectured \cite{AVqdef,superA} to agree with the augmentation polynomial of a knot \cite{NgFramed}. The curve arising as the classical limit of recursions for colored superpolynomials is called the super-A-polynomial \cite{superA,Fuji:2013rra}.

On the other hand, by BPS invariants of a knot we mean the Labastida-Mari{\~n}o-Ooguri-Vafa (LMOV) invariants of a knot \cite{OoguriV,Labastida:2000zp,Labastida:2000yw,Labastida:2001ts}, or certain combinations thereof. 

Our aim is to give exact formulas for a certain class of BPS invariants, as well
as their asymptotic expansions to all orders, using the corresponding
algebraic curve. One motivation of our work is as follows. On one hand, as stated above, various algebraic curves associated to knots arise as classical limits of recursion relations for knot polynomials (Jones, HOMFLY, or superpolynomials) colored by symmetric representations \cite{GL,Hikami_AJ,AVqdef,FGS,superA,FGSS,Nawata}. 
On the other hand, as we will show, one can restrict the defining relations between HOMFLY polynomials and LMOV invariants to the case of symmetric representations only. This implies that recursion relations for knots should encode information about LMOV invariants labeled by symmetric representations, and classical limits of these recursions should still capture some of this information. In this paper we make these statements precise. Our main results are the following. 

\begin{proposition}
\label{prop.main}
\rm{(a)} Fix a knot $K$, a natural number $r$ and an integer $i$. Then the BPS invariants $b_{r,i}$ are given by
\be
x\frac{\partial_x y(x,a)}{y(x,a)} 
= -\frac{1}{2} \sum_{r,i} r^2 b_{r,i} \frac{x^r a^i}{1-x^r a^i},     
\label{xdyy}
\ee
where $y=y(x,a) \in 1 + \BQ[[x,a]]$ is an algebraic function of $(x,a)$
that satisfies a polynomial equation
\be
\mathcal{A}(x,y,a)=0 \,.  \label{Axya}
\ee
\rm{(b)} Explicitly, $x \partial_x y/y$ is an algebraic function of $(x,a)$,
and if $x \partial_x y/y=\sum_{n,m \geq 0} a_{n,m} x^n a^m$, and $\mu(d)$ denotes the M\"obius function, then
\be
b_{r,i} = \frac{2}{r^2}   \sum_{d|r,i} \mu(d) a_{\frac{r}{d},\frac{i}{d}}  
\,.   
\label{b-ri}
\ee
\end{proposition}
The BPS invariants $b_{r,i}$ introduced above are certain combinations of LMOV invariants. In the string theory interpretation they are encoded in holomorphic disk amplitudes or superpotentials \cite{OoguriV,AV-discs,AKV-framing} for D-branes conjecturally associated to knots, and more accurately they could be referred to as classical BPS invariants; however in this paper, unless otherwise stated, we simply call them BPS invariants or degeneracies. The definition of $b_{r,i}$ is given in Section~\ref{sub-total}. The polynomial $\mathcal{A}(x,y,a)$, sometimes referred to as the dual A-polynomial, is defined in Section~\ref{ssec-AJ}. Equation \eqref{xdyy} gives exact formulas for the BPS invariants $b_{r,i}$ for arbitrary $r,i$, as well as asymptotic expansions to all orders of $r$ when $(r,i)$ are along a ray. In particular, one obtains exact formulas
and asymptotic expansions for the BPS invariants $b^\pm_r=b_{r,r \cdot c_\pm}$ along the two extreme rays, that we call extremal BPS invariants. The extremal BPS invariants $b^{\pm}_r$ can also be expressed in terms of coefficients of, respectively, maximal and minimal powers of $a$ in HOMFLY polynomials. The coefficients of these extremal powers are also referred to as top-row and bottom-row HOMFLY polynomials (\emph{rows} refer to the components of the diagram representing HOMFLY homology, see e.g. \cite{Gorsky:2013jxa}).

We call the algebraic curves that encode extremal BPS degeneracies the extremal A-polynomials, and for a knot $K$ we denote them $\mathcal{A}^{\pm}_{K}(x,y)$. We also refer to $\mathcal{A}^{+}_{K}(x,y)$ and $\mathcal{A}^{-}_{K}(x,y)$ respectively as top and bottom A-polynomials. In this work, among the others, we determine extremal A-polynomials for various knots. We also discuss their properties; we note here that in particular they are tempered, i.e. the roots of their face polynomials are roots of unity, which is also referred to as the quantizability condition \cite{superA,Fuji:2013rra} and is the manifestation of the so-called $K_2$ condition \cite{abmodel}. The extremal A-polynomials for twist knots, for a family of $(2,2p+1)$ torus knots, and for various knots with up to 10 crossings are given in Appendix \ref{sec-extremeA}. The extremal A-polynomials for twist knots are also given below. 

Fix an integer $p \neq 0,1$ and consider the family of 
hyperbolic twist knots $K_p$. For $p=-1,-2,-3,\ldots$ these are $4_1,6_1,8_1,\ldots$ knots; for $p=2,3,4\ldots$ these are $5_2,7_2,9_2,\ldots$ knots. Much more can be said for this family of knots.

\begin{proposition}
\label{prop.twist}
The extremal BPS invariants of twist knots are given by
\be
b^-_{K_p,r} =  -\frac{1}{r^2}\sum_{d|r} \mu\big(\frac{r}{d}\big) {3d-1 \choose d-1},\qquad  b^+_{K_p,r} =  \frac{1}{r^2}\sum_{d|r} \mu\big(\frac{r}{d}\big) {(2|p|+1)d - 1  \choose d-1}   
\label{br-neg-intro}
\ee
for $p \leq -1$ and 
\be
b^-_{K_p,r} =  \frac{1}{r^2}\sum_{d|r} \mu\big(\frac{r}{d}\big) (-1)^{d+1} {2d-1 \choose d-1},\qquad   b^+_{K_p,r} =  \frac{1}{r^2}\sum_{d|r} \mu\big(\frac{r}{d}\big) (-1)^d {(2p+2)d - 1  \choose d-1}    \label{br-pos-intro}
\ee
for $p \geq 2$. 
\end{proposition}
Our formulas and the integrality of the BPS invariants
lead to nontrivial integrality statements among sequences of rational numbers.
In particular, it implies that $b^\pm_{K_p,r}$ are integers for all natural numbers $r$ and all integers $p \neq 0,1$.
Note that this implies a nontrivial statement, that for fixed $r$ the sums in expressions (\ref{br-neg-intro}) and (\ref{br-pos-intro}) are divisible by $r^2$. Also note that the above BPS degeneracies are determined explicitly for an infinite range of $r$; this is in contrast to other LMOV invariants in literature, which were determined explicitly only for some finite range of labeling representations (Young diagrams consisting up to several boxes)
\cite{Labastida:2000zp,Labastida:2000yw,Labastida:2001ts,Ramadevi_Sarkar,Zodinmawia:2011oya,Zodinmawia:2012kn}.

What's more, we experimentally discover an Improved Integrality for the BPS invariants, observed by Kontsevich for algebraic curves satisfying the $K_2$ condition  \cite{Kontsevich-K2} (which is the same condition as already mentioned above \cite{abmodel}). 

\begin{conjecture}(Improved Integrality)
\rm{} Given a knot there exist nonzero integers $\gamma^\pm$ such that for any $r
\in \BN$
\be
 \frac{1}{r} \gamma^\pm  b^\pm_r \in \BZ .
\label{eq.improved}
\ee
\end{conjecture}
We checked the above conjecture for twist knots, torus knots and several knots
with up to 10  crossings. The values of $\gamma^{\pm}$ for various knots are given in table \ref{c-minmax-aug}. Note that this integrality conjecture is more general than the integrality
of LMOV invariants \cite{OoguriV,Labastida:2000zp,Labastida:2000yw,Labastida:2001ts}, which only implies integrality of $b_r$. It would be interesting to give a physical interpretation of
this property and of the conjectured integer knot invariants $\gamma^\pm$.

Our next proposition illustrates the special geometry of the extremal 
$A$-polynomials of the twist knots.

\begin{proposition}
\rm{(a)} The extremal A-polynomials of twist knots are are given by
\begin{align}
 \mathcal{A}^-_{K_p}(x,y) &= x-y^4+y^6, & 
\mathcal{A}^+_{K_p}(x,y) &= 1-y^2+x y^{4|p|+2}, & p \leq -1,    
\label{Ap-neg-intro} \\
\mathcal{A}^-_{K_p}(x,y) &=1-y^2- x y^4, &
\mathcal{A}^+_{K_p}(x,y) &= 1-y^2+x y^{4p+4}, & p \geq 2.     
\label{Ap-pos-intro}
\end{align}
\rm{(b)}
The algebraic curves $\mathcal{A}^\pm_{K_p}(x,y^{1/2})$ have a distinguished
solution $y=y(x) \in 1 + x \BZ[[x]]$ such that
\be
\boxed{\text{$y$ and $x y'/y$ are algebraic hypergeometric functions}}
\label{eq.fortunate}
\ee
\end{proposition}
Explicit formulas for those solutions can be found in Section \ref{sec-results}.
Note that if $y$ is algebraic, so is $x y'/y$. The converse nearly holds. 
The next theorem was communicated to us by Kontsevich in the 2011 
Arbeitstagung talk. A proof, using the solution to the Grothendieck-Katz 
conjecture, was written in Kassel-Reutenauer~\cite{KS}.

\begin{theorem}
\label{thm.KS}
If $f \in \BZ[[x]]$ is a formal power series with integer coefficients such
that $g=x f'/f$ is algebraic, then $f$ is algebraic. 
\end{theorem}
With the above notation, the converse holds trivially: if $f$ is algebraic,
so is $g$, without any integrality assumption on the coefficients of $f$.
The integrality hypothesis are required in Theorem~\ref{thm.KS}: if $f=e^x$
then $g=x$. Note finally that the class of algebraic hypergeometric functions 
has been completely classified in \cite{beukers:monodromy, beukers:algebraic}. 
Using this, one can also determine the functions that satisfy 
\eqref{eq.fortunate}. We will not pursue this here.

The special geometry property \eqref{eq.fortunate} does not seem to hold 
for non-twist hyperbolic knots. 

Next, we point out more information encoded in extremal A-polynomials.

\begin{lemma}
\label{prop.disc}
The exponential growth rates of $b^{\pm}_r$ are given by the zeros of $y$-discriminant of the corresponding extremal A-polynomials $\mathcal{A}^{\pm}_{K}(x,y)$
\begin{align}
x_0=\lim_{r\to\infty} \frac{b^{\pm}_r}{b^{\pm}_{r+1}}\quad \Rightarrow\quad 
\disc_y \mathcal{A}_K^{\pm}(x_0,y)=0
\end{align}
\end{lemma}

For example, for $K_p$ torus knots with $p\leq -1$, the $y$-discriminant of $\mathcal{A}^-_{K_p}(x,y)$ in (\ref{Ap-neg-intro}) is
given by 
$$
\disc_y \mathcal{A}^-_{K_p}(x,y)= -64 x^3 (-4 + 27 x)^2
$$
and its zero $x_0=\frac{4}{27}$ matches the exponential growth rate 
of $b^-_{K_p,r}$ in (\ref{br-neg-intro}), which can be obtained from Stirling formula.

Furthermore, we generalize the above analysis of extremal A-polynomials and extremal BPS states by introducing the dependence on variables $a$ and $t$. As conjectured in \cite{AVqdef,superA}, the augmentation polynomials \cite{Ng,NgFramed}, that depend on variable $a$,  should agree with  Q-deformed polynomials obtained as classical limits of recursions for colored HOMFLY polynomials \cite{AVqdef}. We verify that this is indeed the case by computing corresponding BPS invariants from augmentation polynomials, and verifying that they are consistent with LMOV invariants determined from known colored HOMFLY polynomials for various knots with up to 10 crossings. While using our method we can determine BPS invariants for arbitrarily large representations $S^r$ from augmentation polynomials, for more complicated knots the colored HOMFLY polynomials are known explicitly only for several values of $r$, see e.g. \cite{Nawata:2013qpa,Wedrich:2014zua}. Nonetheless this is already quite a non-trivial check.

Finally we introduce the dependence on the parameter $t$, consider refined BPS degeneracies arising from appropriate redefinitions of super-A-polynomials, 
and show their integrality.

We note that apart from the relation to the LMOV invariants our results have an interpretation also from other physical perspectives. In particular, recently a lot of attention has been devoted to the so-called 3d-3d duality, which relates knot invariants to 3-dimensional $\mathcal{N}=2$ theories \cite{DGG,superA,Chung:2014qpa}. In this context the A-polynomial curves, such as (\ref{Axya}) or extremal A-polynomials, represent the moduli space of vacua of the corresponding $\mathcal{N}=2$ theories. Furthermore, the equation (\ref{xdyy}) can be interpreted as imposing (order by order) relations between BPS degeneracies, which should arise from relations in the corresponding chiral rings. We also note that in the corresponding brane system extremal invariants arise from the $\mathbb{C}^3$ limit of the underlying resolved conifold geometry, and the precise way of taking this limit is encoded in the integers $c_{\pm}$ mentioned below (\ref{b-ri}) that specify the extreme rays. We comment on these and other physical interpretations in section \ref{sec-conclude} and plan to analyze them further in future work.


\section{BPS invariants of knots from algebraic curves}    \label{sec-BPS}


\subsection{BPS invariants for knots}    \label{sub-total}

In this section we recall the formulation of Labastida-Mari{\~n}o-Ooguri-Vafa (LMOV) invariants and discuss its form in case of $S^r$-colored HOMFLY polynomials. The starting point is to consider the Ooguri-Vafa generating function \cite{OoguriV,Labastida:2000zp,Labastida:2000yw,Labastida:2001ts}
\be
Z(U,V) = \sum_R  \textrm{Tr}_R U \, \textrm{Tr}_R V = \exp\Big(  \sum_{n=1}^{\infty} \frac{1}{n} \Tr U^n \Tr V^n \Big),
\ee
where $U=P\,\exp\oint_K A$ is the holonomy of $U(N)$ Chern-Simons gauge field along a knot $K$, $V$ can be interpreted as a source, and the sum runs over all representations $R$, i.e. all two-dimensional partitions. The LMOV conjecture states that the expectation value of the above expression takes the following form
\be
\big\langle Z(U,V) \big\rangle = \sum_R P_{K,R}(a,q) \textrm{Tr}_R V  = \exp \Big(  \sum_{n=1}^\infty \sum_R \frac{1}{n} f_{K,R}(a^n,q^n) \textrm{Tr}_R V^n  \Big),    \label{ZUV}
\ee
where the expectation value of the holonomy is identified with the unnormalized HOMFLY polynomial of a knot $K$, $\langle \textrm{Tr}_R U \rangle = P_{K,R}(a,q)$, and the functions $f_{K,R}(a,q)$ take form
\be
f_{K,R}(a,q) = \sum_{i,j} \frac{N_{R,i,j} a^i q^j}{q-q^{-1}},  \label{fR}
\ee
where $N_{R,i,j}$ are famous BPS degeneracies, or LMOV invariants, a term that we will use interchangably; in particular they are conjectured to be integer. In string theory interpretation they count D2-branes ending on D4-branes that wrap a Lagrangian submanifold associated to a given knot $K$. From two-dimensional space-time perspective D2-branes are interpreted as particles with charge $i$, spin $j$, and magnetic charge $R$. For a fixed $R$ there is a finite range of $i$ and $j$ for which $N_{R,i,j}$ are non-zero. 

In what follows we are interested in the case of one-dimensional source $V=x$. In this case $\Tr_R V \neq 0$ only for symmetric representations $R=S^r$ (labeled by partitions with a single row
with $r$ boxes \cite{AM}), so that $\textrm{Tr}_{S^r}(x) = x^r$. For a knot $K$, let us denote $P_K(x,a,q)=\langle Z(U,x) \rangle$, and let $P_{K,r}(a,q) \in \BQ(a,q)$ denote the $S^r$-colored HOMFLY polynomial of $K$. For a detailed 
definition of the latter, see for example \cite{AM}. In this setting (\ref{ZUV}) reduces to the following expression
\be
P_K(x,a,q) = \sum_{r=0}^\infty P_{K,r}(a,q) x^r =   
\exp\Big( \sum_{r,n\geq 1} \frac{1}{n} f_{K,r}(a^n,q^n)x^{n r}\Big).
\label{Pz2}
\ee
Note that we use the unnormalized (or unreduced) HOMFLY polynomials, so that the unknot is normalized as
$P_{{\bf 0_1},1}(a,q)=(a-a^{-1})/(q-q^{-1})$.
Often, we will drop the knot $K$ from the notation.
Note that $f_r(a,q)$ is a universal polynomial (with rational coefficients) of $P_{r/d}(a^d,q^d)$ for all divisors $d$ or $r$. For instance, we have:
\bea
f_1(a,q) &=& P_1(a,q), \nonumber\\
f_2(a,q) &=& P_2(a,q) - \frac{1}{2}P_1(a,q)^2 -\frac{1}{2} P_1(a^2,q^2), \nonumber\\
f_3(a,q) &=& P_3(a,q) - P_1(a,q)P_2(a,q) + \frac{1}{3}P_1(a,q)^3 - \frac{1}{3} P_1(a^3,q^3) \label{f-P}  \\
f_4(a,q) &=& P_4(a,q) - P_1(a,q)P_3(a,q) - \frac{1}{2}P_2(a,q)^2 + P_1(a,q)^2P_2(a,q) + \nonumber\\
& & -\frac{1}{4}P_1(a,q)^4 -\frac{1}{2}P_2(a^2,q^2)+\frac{1}{4}P_1(a^2,q^2)^2. \nonumber
\eea
It follows that $f_r(a,q) \in \BQ(a,q)$. The LMOV conjecture asserts that 
$f_r(a,q)$ can be expressed as a finite sum
$$
f_{r}(a,q) = \sum_{i,j} \frac{N_{r,i,j} a^i q^j}{q-q^{-1}} ,
\qquad
N_{r,i,j} \in \BZ \,.
$$ 
and in this case the BPS degeneracies $N_{r,i,j}$ are labeled by a natural number $r$.

We now explain how to extract BPS degeneracies from the 
generating function \eqref{Pz2}. First we write it in product form
\begin{eqnarray}
P(x,a,q) &=& \sum_r P_r(a,q) x^r 
\exp\Big( \sum_{r,n\geq 1; i,j} \frac{1}{n} 
\frac{N_{r,i,j}(x^r a^i q^j)^n}{q^{n}-q^{-n}}\Big)  \nonumber\\
&=& \exp\Big( \sum_{r\geq 1;i,j;k\geq 0} N_{r,i,j}
\log(1-x^r a^i q^{j+2k+1} )\Big)  \nonumber\\
&=& \prod_{r\geq 1;i,j;k\geq 0} \Big(1 - x^r a^i q^{j+2k+1} \Big)^{N_{r,i,j}} 
\label{Pr-LMOV} \\
&=& \prod_{r\geq 1;i,j} \Big( x^r a^i q^{j+1};q^2 \Big)_\infty^{N_{r,i,j}} 
\nonumber
\end{eqnarray}
where the $q$-Pochhammer symbol (or quantum dilogarithm) notation is used
\be
(x;q)_\infty =\prod_{k=0}^\infty (1-x q^k) \,.    \label{qdilog}
\ee
Then, in the limit $q=e^{\hbar} \to 1$, using well known asymptotic expansion of the quantum dilogarithm (see e.g. \cite{abmodel}), we get the following asymptotic 
expansion
 of $P(x,a,e^{\hbar})$
\begin{eqnarray}
P(x,a,e^{\hbar}) &=& \exp\Big( \sum_{r,i,j} N_{r,i,j} 
\big( \frac{1}{2\hbar}\textrm{Li}_2(x^r a^i) -\frac{j}{2}\log(1-x^r a^i) 
+O(\hbar) \big) \Big)= \nonumber \\
&=&\exp\Big(\frac{1}{2\hbar}\sum_{r,i}  b_{r,i} \textrm{Li}_2(x^r a^i) 
-\sum_{r,i,j} \frac{j}{2}N_{r,i,j}\log(1-x^r a^i) +O(\hbar) \Big)
\label{Px-asympt} \\
&=& \exp\Big(\frac{1}{\hbar} S_0(x,a)+S_1(x,a) + O(\hbar)\Big),   \nonumber
\end{eqnarray}
where 
\begin{align}
S_0(x,a) &= \frac{1}{2} \sum_{r,i}  b_{r,i} \textrm{Li}_2(x^r a^i), \\
S_1(x,a) &= -\frac{1}{2}j\sum_{r,i,j} \log(1-x^r a^i).   
\label{Px-dilog}
\end{align}
Above we introduced 
\be
b_{r,i} = \sum_j N_{r,i,j} \,
\label{eq.btotal}
\ee
that appear at the lowest order in $\hbar$ expansion in the exponent of (\ref{Px-asympt}) and can be interpreted as the classical BPS degeneracies. These degeneracies are of our main concern. In the string theory interpretation they determine holomorphic disk amplitudes or superpotentials \cite{OoguriV,AV-discs,AKV-framing} for D-branes conjecturally associated to knots. In what follows, unless otherwise stated, by BPS degeneracies we mean these numbers.

Our next task is to compute $S_0(x,a)$. To do so, we use a linear $q$-difference
equation for $P(x,a,q)$ reviewed in the next section.


\subsection{Difference equations and algebraic curves}    \label{ssec-AJ}

In this section we introduce various algebraic curves associated to knots.
First, recall that the colored Jones polynomial $J_{K,r}(q) \in \BZ[q^{\pm 1}]$ of a knot $K$ can 
be defined as a specialization of the colored HOMFLY polynomial:
$$
J_{K,r}(q)=P_{K,r}(q^2,q) \,.
$$
It is known that the colored Jones polynomial satisfies a linear $q$-difference
equation of the form
\be
\widehat{A}_K(\hat M, \hat L,q) J_{K,r}(q) = 0, 
\ee
where $\widehat{A}_K$ is a polynomial in all its arguments, and $\hat M$ and $\hat L$ are operators that satisfy the relation $\hat L \hat M = q \hat M \hat L$ and act on colored Jones polynomials by
\be
\hat M J_{K,r}(q) = q^r J_{K,r}(q),\qquad \quad 
\hat L J_{K,r}(q) = J_{K,r+1}(q).    \label{MhatLhat}
\ee
The AJ Conjecture states that
\be
\widehat{A}_K(\hat M, \hat L,1)  = A_K(M,L)
\ee
where $A_K(M,L)$ is the A-polynomial of $K$ \cite{CCGLS}. Likewise, we will
assume that the colored HOMFLY polynomial of a knot satisfies a linear
$q$-difference equation of the form
\be
\widehat{A}(\hat M, \hat L,a,q) P_r(a,q) = 0 \,. \label{Ahat-a}
\ee
The corresponding 3-variable polynomial $A(M,L,a)=A(M,L,a,1)$ defines a family
of algebraic curves parametrized by $a$. A further conjecture \cite{AVqdef,superA} identifies
the 3-variable polynomial $A(M,L,a)$ with the augmentation polynomial of
knot contact homology \cite{Ng,NgFramed}.

We further assume the existence of the super-A-polynomial, i.e. the refined colored HOMFLY polynomial $P_r(a,q,t)$ of a knot, that specializes at $t=-1$ to the usual colored HOMFLY
polynomial, and that also satisfies a linear $q$-difference equation
\cite{superA,FGSS}
\be
\widehat{A}^{\textrm{super}}(\hat M,\hat L, a,q,t) P_r(a,q,t) = 0.      
\ee
The specialization $\widehat{A}^{\textrm{super}}(M,L, a,1,t)$ can be thought
of as an $(a,t)$-family of A-polynomials of a knot.

In the remainder of this section we discuss a dual version $\mathcal{A}(x,y,a)$
of the algebraic curve $A(M,L,a)$. 

\begin{lemma}
Fix a sequence $P_r(a,q)$ which is annihilated by an operator 
$\widehat A (\hat M,\hat L,a,q)$ and consider the
generating function $P(x,a,q)=\sum_{r=0}^{\infty} P_r(a,q) x^r$. Then,  
\be
\widehat{A}(\hat{y},\hat{x}^{-1},a,q) P(x,a,q) = const,  \label{AyxP}
\ee
where
\be
\hat{x} P(x,a,q) = x P(x,a,q),\qquad \quad \hat{y} P(x,a,q) = P(qx,a,q)   \label{lemma-Px}
\ee
satisfy $\hat{x} \hat{y} = q \hat{y} \hat{x}$, and $const$ is a $q$-dependent term that vanishes in the limit $q\to 1$. 
\end{lemma}

\noindent
\emph{Proof.} 
We have
\be
\widehat{A}(\hat M,\hat L) P(x,a,q) = 
\sum_r x^r \widehat{A}(\hat M,\hat L) P_r(a,q) = 0. \label{AhatMLP}
\ee
On the other hand, acting with $\hat M$ and $\hat L$ on this generating function (and taking care of the boundary terms) we get
\bea
\hat M P(x,a,q) &=& \sum_{r=0}^{\infty} P_r(a,q) (qx)^r = P(qx,a,q), \\
\hat L P(x,a,q) &=& \sum_{r=0}^{\infty} P_{r+1}(a,q)x^r = \frac{1}{x} \Big(P(x,a,q) - P_0(a,q) \Big)\nonumber.
\eea
Therefore the action of $\hat M$ and $\hat L$ on $P(x)$ can be identified, respectively, with the action of operators $\hat{y}$ and $\hat{x}^{-1}$, up to the subtlety in the boundary term arising from $r=0$. From the property of the recursion relations for the HOMFLY polynomial the result follows.
\qed

Applying the above lemma to the colored HOMFLY polynomial $P_r(a,q)$, this
motivates us to introduce the operator
\be
\widehat{\mathcal{A}}(\hat{x},\hat{y},a,q) = 
\widehat{A}(\hat{y},\hat{x}^{-1},a,q),
\ee
so that (\ref{AyxP}) can be simply written as
\be
\widehat{\mathcal{A}}(\hat{x},\hat{y},a,q)  P(x,a,q) = const. \label{calAxyP}
\ee
In the limit $q\to 1$ the right hand side vanishes and we can consider the algebraic curve
\be
\mathcal{A}(x,y,a)= A(y,x^{-1},a).  \label{calAxy}
\ee


\subsection{The Lambert transform}

In this section we recall the Lambert transform of two sequences $(a_n)$
and $(b_n)$ which is useful in the proof of Proposition \ref{prop.main}.

\begin{lemma}
\label{lem.1}
\rm{(a)} Consider two sequences $(a_n)$ and $(b_n)$ for $n=1,2,3,\dots$ that
satisfy the relation
\begin{equation}
\label{eq.ab}
a_n = \sum_{d | n} b_d
\end{equation}
for all positive natural numbers $n$. Then we have:
\begin{equation}
\label{eq.ba}
b_n = \sum_{d | n} \mu\left(\frac{n}{d}\right) a_d
\end{equation}
where $\mu$ is the M\"obius function. Moreover, we have the Lambert 
transformation property
\begin{equation}
\label{eq.gf}
\sum_{n=1}^\infty a_n q^n = 
\sum_{n=1}^\infty b_n \frac{q^n}{1-q^n} 
\end{equation} 
and the Dirichlet series property
\begin{equation}
\label{eq.gf2}
\sum_{n=1}^\infty \frac{a_n}{n^s} = \zeta(s)
\sum_{n=1}^\infty \frac{b_n}{n^s} \,.
\end{equation} 
\rm{(b)} If $(a_n)$ has an asymptotic expansion
\begin{equation}
\label{eq.asa}
a_n \sim \sum_{(\lambda,\alpha)} \lambda^n n^{\alpha}
\left( c_0 + \frac{c_1}{n} + \frac{c_2}{n^2}
+ \dots \right)
\end{equation} 
where the first sum is a finite sum of pairs $(\lambda,\alpha)$ such that
$|\lambda|$ is fixed, then so does $(b_n)$ and vice-versa.
\end{lemma}

\noindent
\emph{Proof.}
Part (b) follows from Equation~\eqref{eq.gf2} easily. For a detailed discussion,
see the appendix to \cite{zeidler} by D. Zagier.
\qed

\begin{lemma}
\label{lem.2}
Suppose $y \in \BZ[[x]]$ is algebraic with constant term $y(0)=1$. Write
$$
y=1+\sum_{n=1}^\infty c_n x^n = \prod_{n=1}^\infty (1-x^n)^{b_n}.
$$
If $y$ has a singularity in the interior of the unit circle then $(b_n)$
has an asymptotic expansion of the form~\eqref{eq.asa}. Moreover, the
singularities of the multivalued function $y=y(x)$ are the complex roots of the
discriminant of $p(x,y)$ with respect to $y$, where $p(x,y)=0$ is a 
polynomial equation.
\end{lemma}

\noindent
\emph{Proof.}
We have:
$$
\log y =\sum_{n=1}^\infty b_n \log(1-x^n)
$$
thus if $z=x d\log y=xy'/y$, then we have
$$
z = \sum_{n=1}^\infty n b_n \frac{x^n}{1-x^n} .
$$
Now $z$ is algebraic by the easy converse to Theorem~\ref{thm.KS}. It follows 
that the coefficients $(a_n)$ of its Taylor series
$$
z=\sum_{n=1}^\infty a_n x^n
$$
is a sequence of Nilsson type~\cite{Ga:nilsson}. Since $z$ is algebraic, 
there are no $\log n$ terms in the asymptotic expansion. 
Moreover, the exponential growth rate is bigger than 1, in absolute value. 
Part (b) of Lemma~\ref{lem.1} concludes the proof.
\qed


\subsection{Proof of Proposition 1.1.}

Let us define
\begin{equation}
y(x,a) = 
\lim_{q\to 1} \frac{P(qx,a,q)}{P(x,a,q)} = 
\lim_{q\to 1}  \prod_{r\geq 1;i,j;k\geq 0} 
\Big(\frac{1 - x^r a^i q^{r+j+2k+1} }{1 - x^r a^i q^{j+2k+1}}\Big)^{N_{r,i,j}}  
= \prod_{r\geq 1;i} (1 - x^r a^i)^{-r b_{r,i} / 2}  .
\label{sfx-prod}
\end{equation}
If $P(x,a,q)$ is annihilated by $\widehat{\mathcal{A}}(\hat{x},\hat{y},a,q)$,
it follows that $y=y(a,x)$ is a solution of the polynomial equation
\be
\mathcal{A}(x,y,a) = 0.
\ee
Indeed, divide the recursion 
$$
\widehat{\mathcal{A}}(\hat{x},\hat{y},a,q) P(x,a,q)=0
$$
by $P(x,a,q)$, and observe that
$$
\lim_{q \to 1} \frac{P(q^jx,a,q)}{P(x,a,q)} =
\prod_{l=1}^j \lim_{q \to 1} \frac{P(q^l x,a,q)}{P(q^{l-1}x,a,q)} =
y(x,a)^j.
$$
Taking the logarithm and then differentiating \eqref{sfx-prod} concludes
Equation \eqref{xdyy}. Part (b) of Proposition \ref{prop.main} follows from
Lemma \ref{lem.1}.
\qed


\subsection{Refined BPS invariants}

In this section we discuss refined BPS invariants $N_{r,i,j,k}$. In full generality, we can consider the generating function of superpolynomials $P_r(a,q,t)$; suppose that it has the product structure analogous to (\ref{Pr-LMOV}), however with an additional $t$-dependence
\be
P(x,a,q,t) = \sum_{r=0}^\infty P_r(a,q,t) x^r 
=  \prod_{r\geq 1;i,j,k;n\geq 0} 
\Big(1 - x^r a^i t^j q^{k+2n+1} \Big)^{N_{r,i,j,k}} 
\label{Pr-LMOVref} 
\ee
We conjecture that the refined BPS numbers $N_{r,i,j,k}$ encoded in this expression are integers. As in this work we are mainly interested in invariants in the $q\to 1$ limit, encoded in  (classical) algebraic curves, let us denote them as
\be
b_{r,i,j} = \sum_k N_{r,i,j,k}.
\ee 
In this case the curves in question are of course the dual versions of super-A-polynomial $A^{\rm super}(M,L,a,t)$, with arguments transformed as in (\ref{calAxy}), i.e.
\be
\mathcal{A}(x,y,a,t) = A^{\rm super}(y,x^{-1},a,t)= 0.
\ee
Solving this equation for $y=y(x,a,t)$ and following steps that led to (\ref{xdyy}), we find now
\be
x\frac{\partial_x y(x,a,t)}{y(x,a,t)} = \frac{1}{2} \sum_{r,i,j} r^2 b_{r,i,j} \frac{x^r a^i t^j}{1-x^r a^i t^j},     \label{xdyy-ref}
\ee
and from such an expansion $b_{r,i,j}$ can be determined. Conjecturally these should be integer numbers; as we will see in several examples this turns out to be true. 



\section{Extremal invariants}    \label{sec-extreme}

\subsection{Extremal BPS invariants}

In this section we define extremal BPS invariants of knots. If $P_r(a,q)$ is
a $q$-holonomic sequence, it follows that the minimal and maximal exponent with respect to
$a$ is a quasi-linear function of $r$, for large enough $r$. This follows 
easily from the Lech-Mahler-Skolem theorem as used in \cite{Ga:degree} and is 
also discussed in detail in \cite{vanderveen}. We now restrict our attention to 
knots that satisfy
\be
P_r(a,q) = \sum_{i=r \cdot c_-}^{r\cdot c_+} a^i p_{r,i}(q)      
\label{Pr-minmax}
\ee
for some integers $c_\pm$ and for every natural number $r$, where
$p_{r,r \cdot c_\pm}(q) \neq 0$.
This is a large class of knots -- in particular two-bridge knots and torus knots have this property.
For such knots, we can consider the extremal parts 
of the colored HOMFLY polynomials (i.e. their top and bottom rows, that is the coefficients of maximal and minimal powers of $a$), defined as one-variable polynomials
\be
P^\pm_r(q) = p_{r,r \cdot c_\pm}(q).         \label{Pminmax}
\ee
Likewise, we define the extremal LMOV invariants by 
\be
f^\pm_r(q) = 
\sum_j \frac{N_{r, r \cdot c_\pm,j} q^j}{q - q^{-1}} \,,  \label{f-minmax}
\ee
and the extremal BPS invariants by
\begin{equation}
b^\pm_r = b_{r,r \cdot c_\pm} = \sum_j N_{r,r \cdot c_\pm,j}.         \label{br-minmax}
\end{equation}
We also refer to $b^+_r$ and $b^-_r$ as, respectively, top and bottom BPS invariants.  
Finally, we define the extremal part of the generating series $P(x,a,q)$
by
\be
P^\pm(x,q) = \sum_{r=0}^\infty P^\pm_r(q) x^r 
= \prod_{r\geq 1;j;k\geq 0} 
\Big(1 - x^r q^{j+2k+1} \Big)^{N_{r,r \cdot c_\pm,j}}.
\label{Pr-LMOV-minmax} 
\ee
The analogue of Equation \eqref{Px-asympt} is
\be
P^\pm(x,e^{\hbar}) = 
\exp\Big(\frac{1}{2\hbar}\sum_{r}  b^\pm_r \textrm{Li}_2(x^r) 
-\sum_{r,j} \frac{j}{2}N_{r,r \cdot c_\pm,j}\log(1-x^r) +O(\hbar) \Big). 
\label{Px-asympt-minmax}
\ee

It follows from the LMOV conjecture that $b^{\pm}_r$, as combinations of LMOV invariants $N_{r,r\cdot c_{\pm},j}$, are integer. Moreover, according to the Improved Integrality conjecutre \ref{eq.improved}, for each knot one can find integer numbers $\gamma^{\pm}$, such that 
\be
 \frac{1}{r} \gamma^\pm  b^\pm_r \in \BZ  .
\ee
The numbers $\gamma^{\pm}$ can be regarded as new invariants of a knot. We compute these numbers for various knots  in section \ref{sec-results}, with the results summarized in table \ref{c-minmax-aug}.

\begin{table}
\be
\begin{array}{|c|c|c||c|c|c||c|c|c|}
\hline 
\textrm{\bf Knot} & \  \gamma^- \  & \  \gamma^+ \  & \textrm{\bf Knot} & \  \gamma^- \  & \  \gamma^+ \  & \textrm{\bf Knot} & \  \gamma^- \  & \  \gamma^+ \  \nonumber \\
\hline 
\hline
\  K_{-1-6k}   \  & 2 & 2 & \  K_{2+6k}\    & 6 & 2 & 6_2 & 3 & 30 \\ 
\  K_{-2-6k} \  & 2 & 3 & \  K_{3+6k}\    & 6 & 3 & 6_3 &  6 &  6 \\ 
\  K_{-3-6k}  \  & 2 & 2 & \  K_{4+6k}\    & 6 & 2 & 7_3 & 1 & 10 \\ 
\  K_{-4-6k}   \  & 2 & 1 & \  K_{5+6k}\  & 6 & 1 & 7_5 & 3 & 30 \\ 
\  K_{-5-6k} \  & 2 & 6  & \  K_{6+6k}\   & 6 & 6  & 8_{19} & 7 & 1 \\
\  K_{-6-6k} \  & 2 & 1 & \  K_{7+6k}\   & 6 & 1 & 10_{124} & 2 & 8 \\ 
\hline  
\hline
\ T_{2,2p+1} \ & \ 2p+3 \ &\  2p-1  & \multicolumn{6}{c|}{   }  \\
\hline 
\end{array}
\ee 
\caption{Improved Integrality: values of  $\gamma^-$ and $\gamma^+$ for various knots. For twist knots $K_p$ the range of the subscript is labeled by $k=0,1,2,\ldots$, and $T_{2,2p+1}$ denotes $(2,2p+1)$ torus knot.} \label{c-minmax-aug}
\end{table}


\subsection{Extremal A-polynomials}

It is easy to see that if $P_r(a,q)$ is annihilated by
$\widehat{A}(\hat M, \hat L,a,q)$, then its extremal part $P^\pm_r(q)$ is
annihilated by the operator $\widehat{A}^\pm(\hat M, \hat L,q)$
obtained by multiplying 
$\widehat{A}(\hat M, \hat L,a^{\mp 1},q)$ by $a^{\pm r c_\pm}$ (to make every 
power of $a$ nonnegative), and then setting $a=0$. This allows us to introduce
the extremal analogues of the curve \eqref{calAxy} defined as distinguished, irreducible factors in 
\be
\mathcal{A}(x a^{- c_\pm},y,a)|_{a^{\mp 1} \to 0}       
\label{Amin}  
\ee
that determine the extremal BPS degeneracies. We call these curves extremal A-polynomials and denote them $\mathcal{A}^\pm(x,y)$. We also refer to $\mathcal{A}^+(x,y)$ and  $\mathcal{A}^-(x,y)$ as top and bottom A-polynomials respectively. The extremal A-polynomials that we determine in this work are listed in Appendix \ref{sec-extremeA}.

Among various interesting properties of extremal A-polynomials we note that they are tempered, i.e. the roots of their face polynomials are roots of unity. This is a manifestation of their quantizability and the so-called $K_2$ condition \cite{abmodel, superA,Kontsevich-K2}, and presumably is related to the Improved Integrality of the corresponding extremal BPS states. 


\subsection{Extremal BPS invariants from extremal A-polynomials}

In this section we give the analogue of Proposition \ref{prop.main} for
extremal BPS invariants.

\begin{proposition}
\label{prop.main.extreme}
\rm{(a)} Fix a knot $K$ and a natural number $r$. Then the extremal 
BPS invariants $b^\pm_{r}$ are given by
\be
x\frac{(y^\pm)'(x)}{y^\pm(x)} 
= \frac{1}{2} \sum_{r\geq 1} r^2 b^\pm_{r} \frac{x^r}{1-x^r}\,,  
\label{xdyy-minmax}
\ee
where $y^\pm=y^\pm(x) \in 1 + \BQ[[x]]$ is an algebraic function of $x$
that satisfies a polynomial equation
$$
\mathcal{A}^\pm(x,y^\pm)=0 \,.
$$
\rm{(b)} Explicitly, $x \partial_x y^\pm/y^\pm$ is an algebraic function of 
$x$ and if 
\be
\label{xdyy-an}
x (y^\pm)'(x)/y^\pm(x)=\sum_{n \geq 0} a^\pm_{n} x^n,
\ee 
then
\be
b^\pm_{r} = \frac{2}{r^2}   \sum_{d|r} \mu(d) a^\pm_{\frac{r}{d}}  
\,.   
\label{b-r}
\ee
\end{proposition}

\noindent
\emph{Proof.}
We define
\be
y^\pm(x) = 
\lim_{q\to 1} \frac{P^\pm(qx,q)}{P^\pm(x,q)} = 
\lim_{q\to 1}  \prod_{r\geq 1;i,j;k\geq 0} 
\Big(\frac{1 - x^r q^{r+j+2k+1} }{1 - x^r q^{j+2k+1}}\Big)^{N_{r,r \cdot
c_\pm,j}}  
= \prod_{r\geq 1;i} (1 - x^r )^{-r b^\pm_r / 2} .
\label{sfx-prod-minmax}
\ee
As in the proof of Proposition \ref{prop.main}, it follows that $y^\pm(x)$
satisfies the polynomial equation
$$
\mathcal{A}^\pm(x,y)=0 \,.
$$
This concludes the first part. The second part follows just as in 
Proposition \ref{prop.main}.
\qed


\section{Examples and computations}      
\label{sec-results}

In this section we illustrate the claims and ideas presented earlier in many examples. First of all, LMOV invariants arise from redefinition of unnormalized knot polynomials. Therefore we recall that the unnormalized superpolynomial for the unknot reads \cite{superA}
\be
P_{{\bf 0_1},r}(a,q,t) = (-1)^{\frac{r}{2}}a^{-r}q^{r}t^{-\frac{3r}{2}} \frac{(-a^2t^3;q^2)_{r}}{(q^2;q^2)_{r}} \, ,
\label{Punknot} 
\ee
and the unnormalized HOMFLY polynomial arises from $t=-1$ specialization of this expression. In what follows we often take advantage of the results for normalized superpolynomials $P^{norm}_r(K,a,q,t)$ for various knots $K$, derived in \cite{superA,FGSS,Nawata}. Then the unnormalized superpolynomials that we need from the present perspective differ simply by the unknot contribution
\be
P_{K,r}(a,q,t) = P_{{\bf 0_1},r}(a,q,t) P^{norm}_{K,r}(a,q,t).    \label{Punnorm}
\ee
In general to get colored HOMFLY polynomials one would have to consider the action of a certain differential \cite{DGR,GS}; however for knots considered in this paper, for which superpolynomials are known, HOMFLY polynomials arise from a simple substitution $t=-1$ in the above formulas.

\begin{wrapfigure}{l}{0.4\textwidth}
\begin{center}
\includegraphics[scale=0.3]{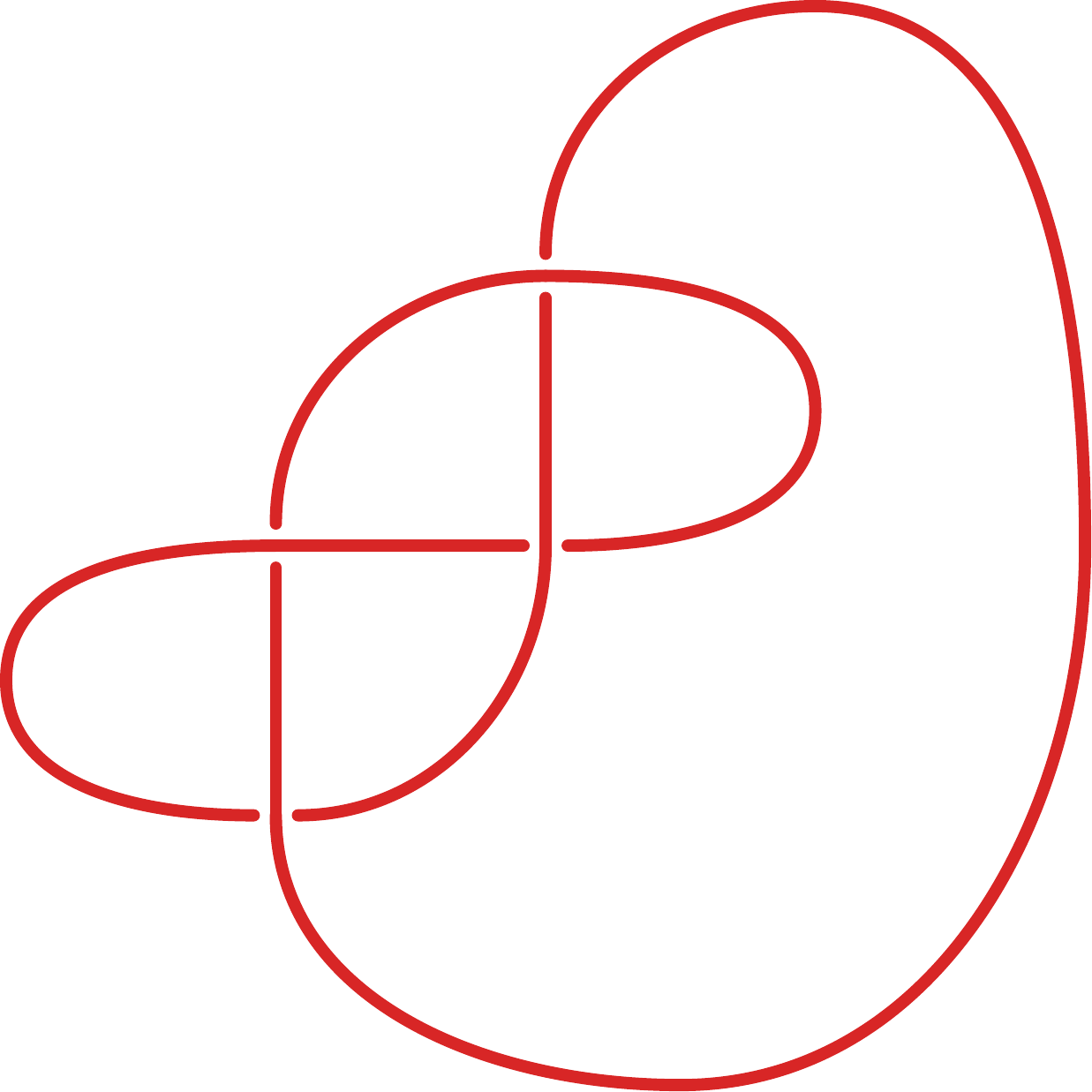}
\end{center}
\caption{The $4_1$ knot.}
\label{fig-41}
\end{wrapfigure}

Let us stress some subtleties related to various variable redefinitions. Super-A-polynomials for various knots, corresponding to the normalized superpolynomials $P^{norm}_r(K,a,q,t)$, were determined in \cite{superA,FGSS,Nawata}; in the current notation we would write those polynomials using variables $M$ and $L$, as $A^{\textrm{super}}(M,L,a,t)$. Super-A-polynomials in the unnormalized case (i.e. encoding asymptotics of unnormalized superpolynomials), which are relevant for our considerations, arise from $A^{\textrm{super}}(M,L,a,t)$ by the substitution
\be
M\mapsto M, \qquad L \mapsto (-a^2t^3)^{-1/2} \frac{1+ a^2 t^3 M}{1-M} L.   \label{MLnorm}
\ee
In what follows we often consider $a$-deformed polynomials, which for knots considered in this paper are again simply obtained by setting $t=-1$ in $A^{\textrm{super}}(M,L,a,t)$. These $a$-deformed polynomials are not yet identical, however closely related (by a simple change of variables) to augmentation polynomials or Q-deformed polynomials; we will present these relations in detail in some examples.


\subsection{The unknot}

Let us illustrate first how our formalism works for the unknot. From the analysis of asymptotics, or recursion relations satisfied by (\ref{Punknot}), the following super-A-polynomial is determined\footnote{More precisely, due to present conventions, one has to substitute $M^2\mapsto x, L\mapsto y, a^2\mapsto a$ to obtain the curve from \cite{superA} on the nose.} \cite{superA}
\be
A(M,L,a,t) = (-a^{-2}t^{-3})^{1/2}(1+a^2 t^3 M^2)-(1-M^2)L.
\ee
The analysis of the refined case essentially is the same as the unrefined, as the dependence on $t$ can be absorbed by a redefinition of $a$. Therefore let us focus on the unrefined case. From (\ref{calAxy}) we find that, up to an irrelevant overall factor, the dual A-polynomial reads
\be
\mathcal{A}(x,y,a) = x-a^2x y^2 -a+ay^2.   \label{Aconi}
\ee
From this expression we can immediately determine
\be
y^2=\frac{1-x a^{-1}}{1-x a},
\ee
and comparing with (\ref{sfx-prod}) we find only two non-zero BPS invariants $b_{1,\pm 1}=\pm 1$ (and from LMOV formulas (\ref{f-P}) one can check that there are also only two non-zero LMOV invariants $N_{1,\pm 1,j}$). These invariants represent of course two open M2-branes wrapping $\mathbb{P}^1$ in the conifold geometry \cite{OoguriV,AV-discs} (for the refined case we also find just two refined BPS invariants). Furthermore, from (\ref{Punknot}) we find $c_{\pm}=\pm 1$, and therefore from (\ref{Amin}) we determine the following extremal A-polynomials
\be
\mathcal{A}^-(x,y) = 1-x-y^2,\qquad \qquad \mathcal{A}^+(x,y) = 1+xy^2-y^2,
\ee
which represent two $\mathbb{C}^3$ limits of the resolved conifold geometry (in terms of $Y=y^2$, the curve (\ref{Aconi}) and the above extremal A-polynomials are the usual B-model curves for the conifold and $\mathbb{C}^3$ respectively).


\subsection{The $4_1$ knot}

As the second example we consider the $4_1$ (figure-8) knot, see figure \ref{fig-41}.
The (normalized) superpolynomial for this knots reads \cite{superA} 
\be
P^{norm}_{r} (a,q,t) = \sum_{k=0}^{\infty} (-1)^k a^{-2k} t^{-2k} q^{-k(k-3)} \frac{(-a^2 t q^{-2},q^2)_k}{(q^2,q^2)_k}  (q^{-2r},q^2)_k (-a^2 t^3 q^{2r}, q^2)_k \,.
\label{Paqt41}
\ee

From this formula, after setting $t=-1$, it is immediate to determine $N_{r,i,j}$ LMOV invariants using the explicit relations (\ref{f-P}), up to some particular value of $r$. Instead, using the knowledge of associated algebraic curves, we will explicitly determine the whole family of these invariants, labeled by arbitrary $r$.

First of all, by considering recursion relations satisfied by $P^{norm}_r$, or from the analysis of its asymptotic behavior for large $r$, the following (normalized) super-A-polynomial is determined\footnote{Again, one has to substitute $M^2\mapsto x, L\mapsto y, a^2\mapsto a$ to obtain the curve from \cite{superA} on the nose.}  in \cite{superA}
\bea
A^{\text{super}} (M,L , a,t) \, &=& \,
a^4 t^5 (M^2-1)^2 M^4 + a^2 t^2 M^4 (1 + a^2 t^3 M^2)^2 L^3 + \label{Asuper41} \\
& & \quad + a^2 t (M^2-1) (1 + t(1-t) M^2  + 2 a^2 t^3(t+1) M^4
    -2 a^2 t^4(t+1) M^6  + \nonumber\\
 & & + a^4 t^6(1-t) M^8  - a^4 t^8 M^{10}) L - (1 + a^2 t^3 M^2) (1 + a^2 t(1-t) M^2  + \nonumber \\
 & & \quad  + 2 a^2 t^2(t+1) M^4  + 2 a^4 t^4(t+1) M^6  + a^4 t^5(t-1) M^8  + a^6 t^7 M^{10}) L^2 . \nonumber
\eea
We will use this formula when we consider refined BPS states; however at this moment let us consider its unrefined (i.e. $t=-1$) version. With the notation
of Section \ref{ssec-AJ} and Equation \eqref{calAxy} we find the dual A-polynomial
\bea
\mathcal{A}(x,y,a) &=& a^3 \left(y^6-y^4\right) +x \left(-a^6 y^{10}+2 a^4 y^8-2 a^2 y^2+1\right)+ \label{calAa-41} \\
& &  + a x^2 \left(a^4 y^{10}-2 a^4 y^8+2 y^2-1\right) +x^3 \left(a^2 y^4-a^4 y^6\right). \nonumber
\eea
From \eqref{Punnorm} and \eqref{Paqt41} we find that the HOMFLY polynomial in the fundamental ($r=1$) representation is given by
\be
P_{1}(a,q) =a^{-3}\frac{q}{\left(1-q^2\right)}   + a^{-1}\frac{q^4+1}{q \left(q^2-1\right)} 
+a \frac{ \left(q^4+1\right)}{q \left(1-q^2\right)} + a^3 \frac{ q}{q^2-1}.
\ee
Comparing this with (\ref{Pr-minmax}) we determine the value of $c_\pm$
\be
c_- = -3, \qquad \quad c_+ =3,   \label{minmax-41}
\ee
so the extremal A-polynomials following from the definition \eqref{Amin} and the result \eqref{calAa-41} are given by
\be
\mathcal{A}^-(x,y) = x - y^4 + y^6,\qquad \quad 
\mathcal{A}^+(x,y) = 1 - y^2 + x y^6.  \label{Aminmax-41}
\ee
Note that $\mathcal{A}^+(x,y^{-1}) = y^{-6} \mathcal{A}^-(x,y)$, i.e. these curves agree up to $y\mapsto y^{-1}$ (and multiplication by an overall monomial 
factor) -- this reflects the fact that the $4_1$ knot is amphicheiral.

We can now extract the extremal BPS invariants from the curves (\ref{Aminmax-41}). As these curves are cubic in terms of $Y=y^2$ variable, we can determine explicit solutions of the corresponding cubic equations. We will use two fortunate coincidences. 

The first coincidence is that the unique solution $Y(x)=1+O(x)$ to the equation
$\mathcal{A}^-(x,Y) = x - Y^2 + Y^3 = 0$ is an algebraic hypergeometric 
function. Explicitly, we have 
\bea
Y(x) = Y^-(x) &=& \frac{1}{3}+\frac{2}{3}\cos\left[\frac{2}{3}\arcsin\left(\sqrt{\frac{3^{3}x}{2^{2}}}\right)\right]  \nonumber\\
&=& \frac{1}{3}\left[1-\sum_{n=0}^{\infty}\frac{2}{3n-1}{3n \choose n}x^{n}\right]   \label{Y-41} \\
&=& \frac{1}{3}+\frac{2}{3}\ _{2}F_{1}\left(-\frac{1}{3},\frac{1}{3};\frac{1}{2};\ \frac{3^{3}x}{2^{2}}\right). \nonumber
\eea
The second coincidence is that $x \partial_x Y^-/Y^-$ is not only algebraic,
but also hypergeometric. Explicitly, we have:
\bea
x\frac{\partial_x Y^-(x)}{Y^-(x)} &=& \frac{1}{3}-\frac{2}{3}\frac{\cos\left[\frac{1}{6}\arccos\left(1-\frac{27x}{2}\right)\right]}{\sqrt{4-27x}} \nonumber \\
&=& - \sum_{n=1}^{\infty} {3n-1 \choose n-1} x^n  \\
&=& \frac{1}{3}-\frac{1}{3}\ _{2}F_{1}\left(\frac{1}{3},\frac{2}{3};\frac{1}{2};\ \frac{3^{3}x}{2^{2}}\right).
\eea
Recalling \eqref{b-r}, we find that 
\be
x\frac{\partial_x Y^-(x)}{Y^-(x)} =\sum_{n=1}^\infty a^-_n x^n, \qquad
a^-_n = -\frac{1}{2} {3n-1 \choose n-1}    \label{an-41}
\ee
so that the extremal bottom BPS degeneracies \eqref{b-r} are given by
\be
b^-_r =  -\frac{1}{r^2}\sum_{d|r} \mu\big(\frac{r}{d}\big) {3d-1 \choose d-1}.   \label{br-41}
\ee
Several values of $b^-_r$ are given in table \ref{br-41-tab}. Note that the integrality of $b^-_r$ implies a nontrivial statement, that for each $r$ the sum in (\ref{br-41}) must be divisible by $r^2$.

\begin{table}
\be
\begin{array}{|c|c|c|}
\hline 
r & b^-_r = -b^+_r & 2\frac{b^-_r}{r}  \nonumber \\
\hline 
1 & -1 &  -2\\
2 &  -1 &  -1 \\
3 &  -3 &  -2 \\
4 &  -10 & -5 \\
5 & -40 &  -16 \\
6 & -171 &  -57 \\
7 & -791 & -226  \\
8 & -3\, 828 & - 957 \\
9 & -19\, 287 &  -4286 \\
10 & -100\, 140 & - 20\, 028\\
11 & -533\, 159 &  -96\, 938 \\
12 & -2\, 897\, 358 & -482\, 893 \\
13 & -16\, 020\, 563 & -2\, 464\, 702  \\
14 &  -89\, 898\, 151 &  -12\, 842\, 593 \\
\ 15 \ &\  -510\, 914\, 700 \ &\ -68\, 121\, 960 \ \\
\hline  
\end{array}
\ee  
\caption{Extremal BPS invariants and their Improved Integrality for 
the $4_1$ knot. 
}      \label{br-41-tab}
\end{table}

In an analogous way we determine top BPS invariants. 
The above mentioned two coincidences persist. The solution 
$Y^+(x)=1/Y^-(x)=1+O(x)$ of the equation $\mathcal{A}^+(x,Y)=1-Y+xY^3=0$ is
given by
\be
Y^+(x) = \frac{2}{\sqrt{3x}}\sin\left[\frac{1}{3}\arcsin\left(\sqrt{\frac{3^{3}x}{2^{2}}}\right)\right] = \sum_{n=0}^{\infty} \frac{x^n}{2n+1} {3n \choose n} = \, _{2}F_{1}\left(\frac{1}{3},\frac{2}{3};\frac{3}{2};\ \frac{3^{3}x}{2^{2}}\right)
\ee
so that
\be
x\frac{\partial_x Y^+(x)}{Y^+(x)} = - x\frac{\partial_x Y^-(x)}{Y^-(x)}.
\ee
Therefore $a_n^+=-a^-_n$ and $b^+_r = -b^-_r$, which is a manifestation of the amphicheirality of the $4_1$ knot. The above results illustrate Proposition \ref{prop.twist} for the $4_1=K_{-1}$
twist knot. 

Experimentally, it also appears that the Improved Integrality holds 
\eqref{eq.improved} with $\gamma^\pm =2$; see table \ref{br-41-tab}.

Next, we discuss the asymptotics of $b^\pm_r$ for large $r$. Stirling's 
formula gives the asymptotics of $a^-_r$, and part (b) of Lemma \ref{lem.1}
concludes that the asymptotics of $b^-_r$ are given by
\begin{align*}
b^-_r &= -\frac{1}{28 \sqrt{3 \pi}} 
\left(\frac{27}{4}\right)^r r^{-1/2} \Big( 
1
-\frac{7}{72 r}
+\frac{49}{10368 r^2}
+\frac{6425}{2239488 r^3}
-\frac{187103}{644972544 r^4}
+O\left(\frac{1}{r}\right)^5 \Big).
\end{align*}
Note that the $Y$-discriminant of $\mathcal{A}^-(x,Y)$ is
given by 
$$
\disc_Y \mathcal{A}^-(x,Y)= -x (-4 + 27 x)
$$
and its root $x=\frac{4}{27}$ matches the exponential growth rate 
of $b^-_r$, as asserted in Lemma \ref{lem.2}.

Finally, we discuss all BPS invariants $b_{r,i}$, not just the extremal ones, i.e. we turn on the
$a$-deformation. To this end it is useful to rescale the variable $x$ in 
\eqref{calAa-41} by $c_-=-3$, so that
\bea
\mathcal{A}(a^3 x,y,a) &=& (x-y^4+y^6) -2 a x y^2 +a^2 \left(2 x^2 y^2-x^2+2 x y^8\right) + \\
&&  -a^3 x y^{10}  +a^4 \left(x^3 y^4+x^2 y^{10}-2 x^2 y^8\right)  -a^5 x^3 y^6  \nonumber
\eea
contains $\mathcal{A}^-(x,y)$ at its lowest order in $a$. Then, from (\ref{xdyy}) and \eqref{b-ri} we can determine the invariants $b_{r,i}$ for this curve; we list some of them in table \ref{aBPS-41-tab}, whose first column of course agrees with the extremal BPS invariants $b^-_r$
given by \eqref{br-41}.

\begin{table}
\begin{small}
\be
\begin{array}{|c|ccccccccc|} \hline
r \setminus i & 0 & 1 & 2 & 3 & 4 & 5 & 6 & 7 & 8 \\ \hline
1 & -1 & 1 & 1 & -2 & 1 & 0 & 0 & 0 & 0 \\
2 & -1 & 2 & 1 & -4 & 3 & -6 & 11 & -8 & 2 \\
3 & -3 & 7 & 2 & -18 & 21 & -23 & 34 & -48 & 82 \\
4 & -10 & 30 & -2 & -88 & 134 & -122 & 150 & -234 & 384 \\
5 & -40 & 143 & -55 & -451 & 889 & -797 & 664 & -978 & 1716 \\
6 & -171 & 728 & -525 & -2346 & 5944 & -5822 & 3134 & -2862 & 6196 \\
7 & -791 & 3876 & -4080 & -12172 & 39751 & -44657 & 17210 & 4958 & 4071 \\
8 & -3828 & 21318 & -29562 & -62016 & 264684 & -347256 & 121276 & 191744 & -263282 \\
9 & -19287 & 120175 & -206701 & -303910 & 1751401 & -2692471 & 1053774 & 2299115 & -4105859 \\
10 & -100140 & 690690 & -1418417 & -1381380 & 11503987 & -20672858 & 10012945 & 21567000 & -46462399 \\ \hline
\end{array}
\nonumber
\ee
\caption{BPS invariants $b_{r,i}$ for the $4_1$ knot.} 
\label{aBPS-41-tab}
\end{small}
\end{table}


\subsection{$5_2$ knot and Catalan numbers}

\begin{wrapfigure}{l}{0.4\textwidth}
\begin{center}
\includegraphics[scale=0.3]{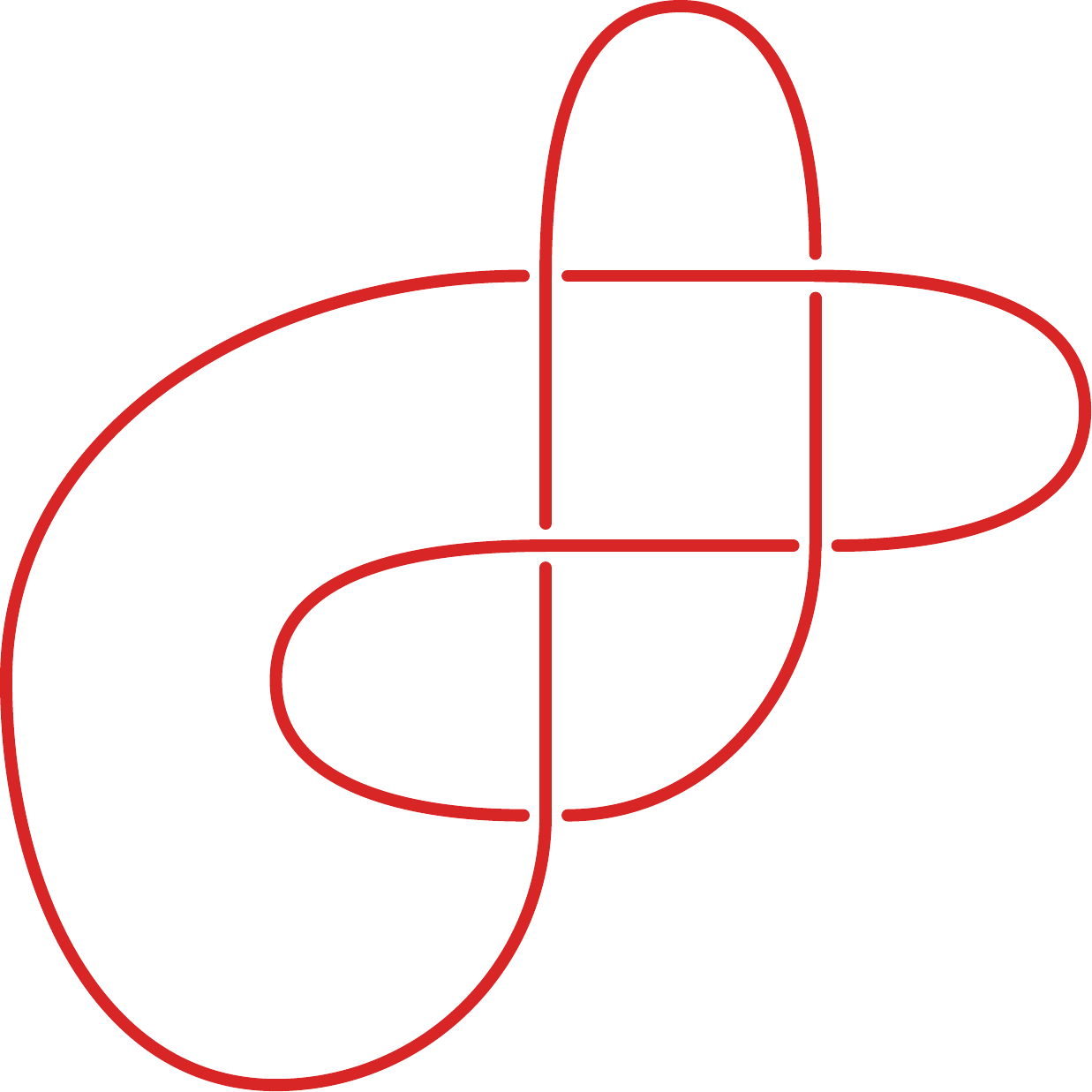}
\end{center}
\caption{The $5_2$ knot.}
\label{fig-52}
\end{wrapfigure}

We can analyze the $K_2=5_2$ knot, see figure \ref{fig-52}, similarly as we did for the figure-8 knot. Starting from the super-A-polynomial derived in \cite{FGSS} and performing redefinitions discussed above, we get the following $a$-deformed algebraic curve (dual A-polynomial)
\bea
\mathcal{A}(x,y,a) &=& a^{10} x^2 y^{16}-2 a^9 x^3 y^{16}+a^8 x^2 y^{12} \left(x^2 y^4-3 y^2-4\right)+a^7 x y^{12} \left(x^2 \left(4 y^2+3\right)-1\right)+ \nonumber \\
&& + a^6 x^2 y^8 \left(-x^2 y^6+5 y^4+3 y^2+6\right)+a^5 x y^{10} \left(2-x^2 \left(3 y^2+2\right)\right) + \label{calAa-52} \\
& & -a^4 x^2 \left(3 y^6+4 y^4-3 y^2+4\right) y^4-a^3 x \left(y^2-1\right) y^4 \left(x^2 \left(y^2-1\right)+y^2+3\right)+\nonumber \\
& & + a^2 x^2 \left(-3 y^6+5 y^4-3 y^2+1\right)+a x \left(-3 y^4+4 y^2-2\right)-y^2+1. \nonumber
\eea
The unnormalized HOMFLY polynomial is given by
\be
P_{1}(a,q) = a \frac{\left(q^4-q^2+1\right)}{q \left(1-q^2\right) } + a^5\frac{ \left(q^4+1\right)}{q \left(q^2-1\right)}+a^7\frac{q}{1-q^2},
\ee
so that 
\be
c^- =1,\qquad\quad c^+=7.
\ee
It follows that the extremal A-polynomials of \eqref{Amin} are given by
\be
\mathcal{A}^-(x,y) = 
1 - y^2 - x y^4 ,\qquad \quad \mathcal{A}^+(x,y) = 1 - y^2 - x y^{12}.  \label{Aminmax-52}
\ee

The two fortunate coincidences of the $4_1$ knot persist for the $5_2$ knot
as well. It is again convenient to use a rescaled variable $Y=y^2$. 
In particular note that the curve $\mathcal{A}^-(x,y)$,   
presented as $1-y^2-xy^4 = 1-Y-x Y^2$, is the curve that encodes 
the Catalan numbers. The latter are the coefficients in the series expansion
\be
Y(x)=Y^-(x) = \frac{-1+\sqrt{1+4x}}{2x} = \sum_{n=0}^{\infty} \frac{1}{n+1}{2n \choose n} (-x)^n. \label{Ymin-52}
\ee
Therefore we have found a new role of Catalan numbers -- they encode BPS numbers for $5_2$ knot (and as we will see, also for other twist knots $K_p$ for $p>1$).

Now we get
\be
x\frac{\partial_x Y^-(x)}{Y^-(x)} = -\frac{1}{2}+\frac{1}{2}\frac{1}{\sqrt{1+4x}} = \sum_{n=1}^{\infty} {2n-1 \choose n-1} (-x)^n = -\frac{1}{2} + \frac{1}{2}\, _1 F_0\left(\frac{1}{2};\ -\frac{2^{2}x}{1^{1}}\right),
\ee
so that
\be
b^-_r =  \frac{1}{r^2}\sum_{d|r} \mu\big(\frac{r}{d}\big) (-1)^{d+1} {2d-1 \choose d-1}.   \label{br-52min}
\ee
Several values of $b^-_r$ are given in table \ref{br-52-tab}.

\begin{table}
\be
\begin{array}{|c|c|c|c|c|}
\hline 
r & b^-_r & b^+_r & 6\frac{b^-_r}{r} & 2\frac{b^+_r}{r}  \nonumber \\
\hline 
1 & -1 & -1 & -6 & -2 \\
2 &  1 &  3& 3 & 3 \\
3 & -1 & -15 & -2 & -10 \\
4 &  2 & 110 & 3 & 55 \\
5 & -5 & -950 & -6 & -380 \\
6 & 13 & 9021 & 13 & 3007 \\
7 & -35 & -91\, 763 & -30 & -26\, 218\\
8 & 100 & 982\, 652 & 75 & 245\, 663 \\
9 & -300 & -10\, 942\, 254 & -200 & -2\, 431\, 612\\
10 & 925 & 125\, 656\, 950 & 555 &25\, 131\, 390\\
11 & -2915 & -1\, 479\, 452\, 887 & -1590 & -268\, 991\, 434\\
12 & 9386 & 17\, 781\, 576\, 786 & 4693 &2\, 963\, 596\, 131\\
13 & -30\, 771& -217\, 451\, 355\, 316 & -14\, 202 & -33\, 454\, 054\, 664\\
14 & 102\, 347 & 2\, 698\, 753\, 797\, 201& 43\, 863 & 385\, 536\, 256\, 743\\
\ 15 \ &\ -344\, 705 \ &\ -33\, 922\, 721\, 455\, 050\ &\ -137\, 882 \ &\  -4\, 523\, 029\, 527\,340\ \\
\hline  
\end{array}
\ee  
\caption{Extremal BPS invariants and their Improved Integrality for $5_2$ knot.} \label{br-52-tab}
\end{table}

In an analogous way for $\mathcal{A}^+(x,Y) = 1 - Y - x Y^{6}$ we get more involved solution
\be
Y(x)^+ = 1 + \sum_{n=1}^{\infty} \frac{1}{n} {6n \choose n-1} (-x)^n = 
_{5}F_{4}\left(\frac{1}{6},\frac{2}{6},\frac{3}{6},\frac{4}{6},\frac{5}{6};\ \frac{6}{5},\frac{4}{5},\frac{3}{5},\frac{2}{5};\ -\frac{6^{6}x}{5^{5}}\right),  \label{Ymax-52}
\ee
so that
\be
x\frac{\partial_x Y^+(x)}{Y^+(x)} = \sum_{n=1}^{\infty} {6n-1 \choose n-1} (-x)^n = -\frac{1}{6}+\frac{1}{6} \, _{5}F_{4}\left(\frac{1}{6},\frac{2}{6},\frac{3}{6},\frac{4}{6},\frac{5}{6};\ \frac{4}{5},\frac{3}{5},\frac{2}{5},\frac{1}{5};\ -\frac{6^{6}x}{5^{5}}\right),
\ee
and in consequence
\be
b_r^+ = \frac{1}{r^2}\sum_{d|r} \mu\big(\frac{r}{d}\big) (-1)^{d} {6d-1 \choose d-1}. \label{br-52max}
\ee
Several values of $b^+_r$ are also given in table \ref{br-52-tab}.

Our discussion illustrates Proposition \ref{prop.twist} for the $5_2=K_2$
twist knot. Experimentally -- see table \ref{br-52-tab} -- it appears that the Improved Integrality \eqref{eq.improved}
holds with
\be
\gamma^- = 6,\qquad\quad \gamma^+=2.
\ee

Next, we consider the asymptotics of the extremal BPS numbers. Using Lemma \ref{lem.1} and Stirling formula for the binomials in \eqref{br-52min} and \eqref{br-52max} we find respectively
\be
\lim_{r\to\infty} \frac{b^-_r}{b^-_{r+1}} = -\frac{1}{4},\qquad\quad \lim_{r\to\infty} \frac{b^+_r}{b^+_{r+1}} = -\frac{5^5}{6^6}.   \label{br-asympt-52}
\ee
This matches with Lemma \ref{lem.2}, since the $Y$-discriminants of 
$\mathcal{A}^\pm(x,Y)$ are given by 
\begin{eqnarray}
\disc_Y \mathcal{A}^-(x,Y) &=& 1 + 4 x,\nonumber \\
\disc_Y \mathcal{A}^+(x,Y) &=& x^4(5^5 + 6^6x).\nonumber
\end{eqnarray}

Finally, we also consider the $a$-deformation of the extremal curves.
Rescaling $x\mapsto a^{-c_-} x$ in \eqref{calAa-52} with $c_-=1$ we get a curve 
that contains $\mathcal{A}^-(x,y)$ at its lowest order in $a$. Then, from 
\eqref{xdyy} and \eqref{b-ri}, the find integral invariants $b_{r,i}$ given in 
table \ref{aBPS-52-tab}. The first column of table \ref{aBPS-52-tab} agrees 
with the values of \eqref{br-52min}.

\begin{table}
\begin{small}
\be
\begin{array}{|c|ccccccccc|} \hline
r \setminus i & 0 & 1 & 2 & 3 & 4 & 5 & 6 & 7 & 8 \\ \hline
1 & -1 & 0 & 2 & -1 & 0 & 0 & 0 & 0 & 0 \\
2 & 1 & -2 & 1 & -4 & 11 & -10 & 3 & 0 & 0 \\
3 & -1 & 4 & 0 & -12 & 23 & -71 & 154 & -162 & 80 \\
4 & 2 & -12 & 14 & 8 & 40 & -226 & 594 & -1542 & 2944 \\
5 & -5 & 36 & -66 & -30 & 132 & -184 & 1550 & -6108 & 16525 \\
6 & 13 & -114 & 302 & -94 & -419 & -660 & 3387 & -12042 & 56209 \\
7 & -35 & 372 & -1296 & 1168 & 1843 & -1400 & -1372 & -27398 & 135350 \\
8 & 100 & -1244 & 5382 & -8014 & -4222 & 12390 & 23462 & -42626 & 163928 \\
9 & -300 & 4240 & -21932 & 45383 & -6044 & -79628 & -31246 & 116368 & 589812 \\
10 & 925 & -14676 & 88457 & -234124 & 160251 & 359514 & -192014 & -1022096 & -86854 \\ \hline
\end{array}
\nonumber
\ee
\caption{BPS invariants $b_{r,i}$ for the $5_2$ knot.} 
\label{aBPS-52-tab}
\end{small}
\end{table}


\subsection{Twist knots}

The $4_1$ and $5_2$ knots are special cases, corresponding respectively to $p=-1$ and $p=2$, in a series of twist knots $K_p$, labeled by an integer $p$. Apart from a special case $p=0$ which is the unknot and $p=1$ which is the
trefoil knot $3_1$, all other twist knots are hyperbolic. In this section we analyze their BPS invariants. The two coincidences of the $4_1$
knot persist for all twist knots. The formulas for $p>1$ are somewhat
different from those for $p<0$, so we analyze them separately. 

We start with $p<0$. In this case the bottom A-polynomial turns out to be the same for all $p$, however the top A-polynomial depends on $p$,  
\be
\boxed{  \mathcal{A}^-_{K_p}(x,y)=x-y^4+y^6,\qquad\quad \mathcal{A}^+_{K_p}(x,y) = 1-y^2+x y^{4|p|+2}  }
\ee
For $p=-1$ these curves of course reduce to those for the $4_1$ knot (\ref{Aminmax-41}). For all $p<0$ the bottom BPS invariants are the same as for the $4_1$ knot \eqref{br-41}
\be
b^-_{K_p,r} =  -\frac{1}{r^2}\sum_{d|r} \mu\big(\frac{r}{d}\big) {3d-1 \choose d-1}, \qquad p <0.   \label{br-min-Kp<0}
\ee
To get top invariants it is convenient to introduce $Y=y^2$ and consider the equation $\mathcal{A}^+_{K_p}(x,Y) = 1-y^2+x Y^{2|p|+1} = 0$, whose solution of interest reads
\bea
Y(x) &=& \sum_{n=0}^{\infty} \frac{1}{2|p|n+1} {(2|p|+1)n \choose n} x^n = \\
&=& _{2|p|}F_{2|p|-1}\left(\frac{1}{2|p|+1},...,\frac{2|p|}{2|p|+1};\frac{2|p|+1}{2|p|},\frac{2|p|-1}{2|p|},\frac{2|p|-2}{2|p|},...,\frac{2}{2|p|};\ \frac{\left(2|p|+1\right)^{2|p|+1}x}{\left(2|p|\right)^{2|p|}}\right) \nonumber
\eea
where $_{2|p|}F_{2|p|-1}$ is the generalized hypergeometric function. Then
\bea
x\frac{Y'(x)}{Y(x)} &=& \sum_{n=1}^{\infty} {(2|p|+1)n - 1  \choose n-1} x^n = \\
&=& \frac{1}{2|p|+1}\left( _{2|p|}F_{2|p|-1}\left(\frac{1}{2|p|+1},...,\frac{2|p|}{2|p|+1};\ \frac{2|p|-1}{2|p|},...,\frac{1}{2|p|};\ \frac{(2|p|+1)^{2|p|+1} x}{\left(2|p|\right)^{2|p|}}\right)  -1\right)\nonumber
\eea
From this expression we find top BPS invariants
\be
b^+_{K_p,r}=  \frac{1}{r^2}\sum_{d|r} \mu\big(\frac{r}{d}\big) {(2|p|+1)d - 1  \choose d-1}, \qquad p<0.   \label{br-max-Kp<0}
\ee

Next, we consider the case $p>1$. Their bottom A-polynomials are the same for all $p$, however top A-polynomials depend on $p$,  
\be
\boxed{ \mathcal{A}^-_{K_p}(x,y)=1-y^2- x y^4,\qquad\quad \mathcal{A}^+_{K_p}(x,y) = 1-y^2+x y^{4p+4} }
\ee
For $p=2$ these curves reduce to the results for $5_2$ knot (\ref{Aminmax-52}). For all $p>1$ the bottom BPS invariants are the same as for $5_2$ knot (\ref{br-52min})
\be
b^-_{K_p,r} =  \frac{1}{r^2}\sum_{d|r} \mu\big(\frac{r}{d}\big) (-1)^{d+1} {2d-1 \choose d-1}, \qquad p>1   \label{br-min-Kp>0}
\ee
which means that Catalan numbers encode these BPS invariants for all $K_{p>1}$ twist knots.

To get top invariants we again introduce $Y=y^2$ and consider the equation $\mathcal{A}^+_{K_p}(x,Y) = 1-Y-x Y^{2p+2} = 0$, whose solution of interest reads
\bea
Y(x) &=& 1 + \sum_{n=1}^{\infty} \frac{1}{n} {(2p+2)n \choose n-1} (-x)^n = \\
&=& _{2p+1}F_{2p}\left(\frac{1}{2p+2},...,\frac{2p+1}{2p+2};\frac{2p+2}{2p+1},\frac{2p}{2p+1},\frac{2p-1}{2p+1},...,\frac{2}{2p+1};-\frac{\left(2p+2\right)^{2p+2}x}{\left(2p+1\right)^{2p+1}}\right) \nonumber
\eea
Then
\bea
x\frac{Y'(x)}{Y(x)} &=& \sum_{n=1}^{\infty} {(2p+2)n - 1  \choose n-1} (-x)^n = \\
&=& \frac{1}{2p+2}\left( _{2p+1}F_{2p}\left(\frac{1}{2p+2},...,\frac{2p+1}{2p+2};\frac{2p}{2p+1},...,\frac{1}{2p+1};-\frac{\left(2p+2\right)^{2p+2}x}{\left(2p+1\right)^{2p+1}}\right)  -1\right) \nonumber
\eea
From this expression we find top BPS invariants
\be
b^+_{K_p,r} =  \frac{1}{r^2}\sum_{d|r} \mu\big(\frac{r}{d}\big) (-1)^d {(2p+2)d - 1  \choose d-1}, \qquad p>1.   
\label{br-max-Kp>0}
\ee

The Improved Integrality holds for twist knots $K_p$ and the values of $\gamma^\pm$ are given in table \ref{c-minmax-aug}. Note that these invariants repeat periodically with period 6 for both positive and negative $p$.

Finally, turning on $a$-deformation also leads to integral invariants $b_{r,i}$, as an example see the results for $6_1= K_{-2}$ knot in table \ref{aBPS-61-tab}.

\begin{table}
\begin{small}
\be
\begin{array}{|c|ccccccccc|} \hline
r \setminus i & 0 & 1 & 2 & 3 & 4 & 5 & 6 & 7 & 8 \\ \hline
1 & -1 & 1 & 1 & -2 & 1 & 0 & 0 & 0 & 0 \\
2 & -1 & 2 & 1 & -4 & 3 & -6 & 11 & -8 & 2 \\
3 & -3 & 7 & 2 & -18 & 21 & -23 & 34 & -48 & 82 \\
4 & -10 & 30 & -2 & -88 & 134 & -122 & 150 & -234 & 384 \\
5 & -40 & 143 & -55 & -451 & 889 & -797 & 664 & -978 & 1716 \\
6 & -171 & 728 & -525 & -2346 & 5944 & -5822 & 3134 & -2862 & 6196 \\
7 & -791 & 3876 & -4080 & -12172 & 39751 & -44657 & 17210 & 4958 & 4071 \\
8 & -3828 & 21318 & -29562 & -62016 & 264684 & -347256 & 121276 & 191744 & -263282 \\
9 & -19287 & 120175 & -206701 & -303910 & 1751401 & -2692471 & 1053774 & 2299115 & -4105859 \\
10 & -100140 & 690690 & -1418417 & -1381380 & 11503987 & -20672858 & 10012945 & 21567000 & -46462399 \\ \hline
\end{array}
\nonumber
\ee
\caption{BPS invariants $b_{r,i}$ for the $6_1$ knot.} 
\label{aBPS-61-tab}
\end{small}
\end{table}


\subsection{Torus knots}

\begin{wrapfigure}{l}{0.4\textwidth}
\begin{center}
\includegraphics[scale=0.3]{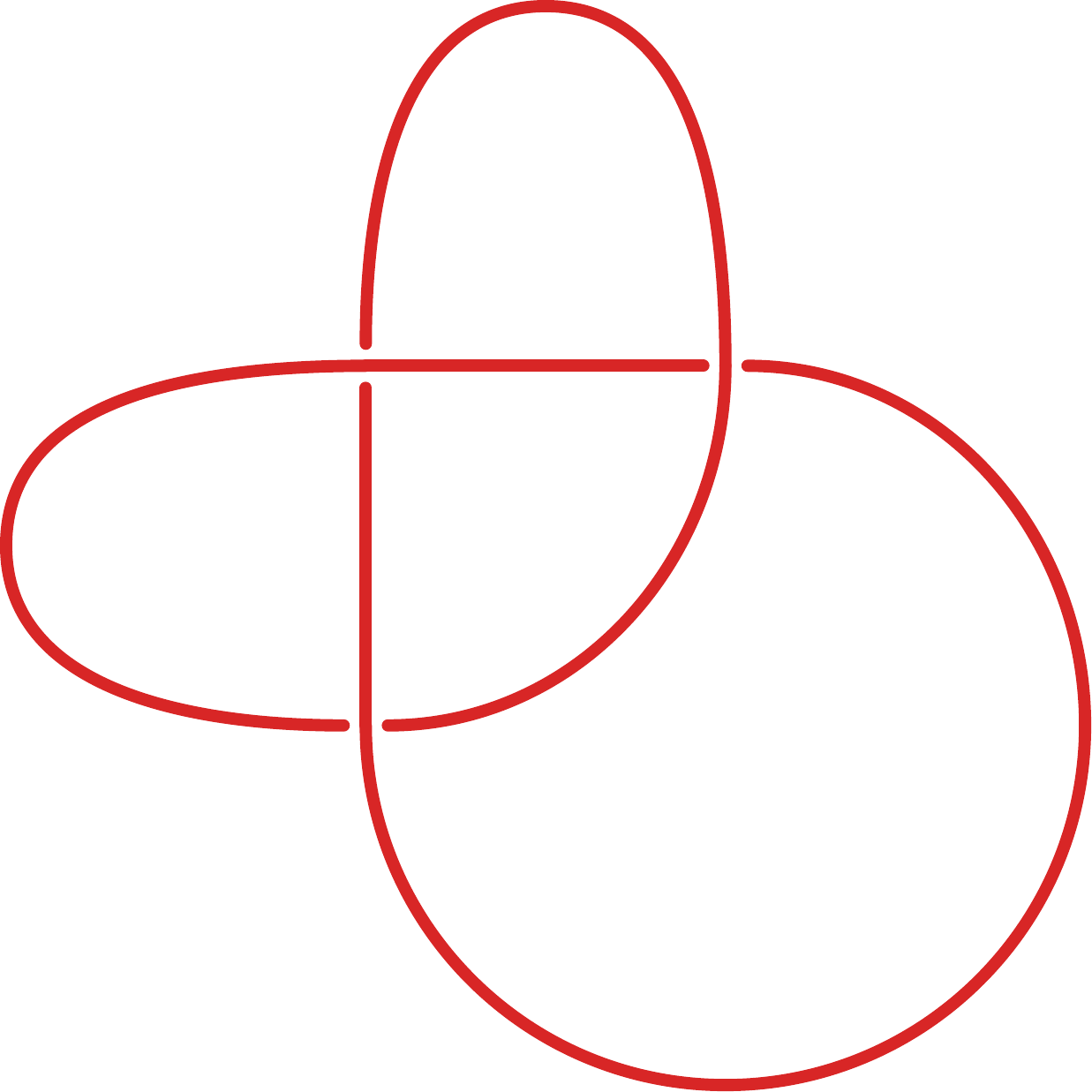}
\end{center}
\caption{The $3_1$ knot.}
\label{fig-31}
\end{wrapfigure}

We can analyze BPS degeneracies for torus knots in the same way as we did for twist knots.  Let us focus on the series of $(2,2p+1)\equiv (2p+1)_1$ knots and present several examples. For the trefoil, $3_1$, see figure \ref{fig-31}, the ($a$-deformed) dual A-polynomial is given by
\be
\mathcal{A}(x,y,a) = -1 + y^2 + a^2 x^2 y^6 + a^5 x y^8  + 
 a x(1 -  y^2 + 2  y^4) - a^3 x y^4(2  + y^2) - a^4 x^2 y^8.
\ee
From the form of the HOMFLY polynomial we find $c_-=1$, so after appropriate rescaling of the above result we get
\be
\mathcal{A}(a^{-1}x,y,a) = (-1 + x + y^2 - x y^2 + 2 x y^4 + x^2 y^6)  - a^2 x y^4 (2 + y^2 + x  y^4)  +a^4 x y^8 .
\ee
The lowest term in $a$ in this expression represents the extremal A-polynomial
\be
\mathcal{A}^-(x,y) = -1 + x + y^2 - x y^2 + 2 x y^4 + x^2 y^6.
\ee
The corresponding BPS invariants are given in table \ref{aBPS-31-tab}.

For the $5_1$ knot we find that $c_-=3$ and the rescaled dual A-polynomial takes form
\bea
\mathcal{A}(a^{-3}x,y,a) &=&     -1 + x + y^2 + 2 x^2 y^{10} + x^3 y^{20}  +  \\
&& +  x y^2 (-1 + 2 y^2) + x y^6 (-2 + 3 y^2) + x^2 y^{12} (-1 + 3 y^2) \nonumber \\
&& + a^2 \left(-x^3 y^{22}-2 x^2 y^{16}-x^2 \left(4 y^2+1\right) y^{12}-x y^{10}-2 x y^4+2 x \left(1-2 y^2\right) y^6\right) + \nonumber\\
&& + a^4 \left(x^2 y^{14}+2 x^2 \left(y^2+1\right) y^{16}+x y^8+x \left(2 y^2-1\right) y^{10}\right) +\nonumber \\
&& + a^6 (-2 x^2 y^{18} - x^2 y^{20}) + a^8 x^2 y^{22} .  \nonumber
\eea
The terms in the first two lines above constitute the extremal A-polynomial $\mathcal{A}^-(x,y)$. Several corresponding BPS invariants are given in table \ref{aBPS-51-tab}.

For the $7_1$ knot the $a$-deformed  A-polynomial takes form of a quite lengthy expression, so we list just some BPS invariants -- with $a$-deformation taken into account -- in the table \ref{aBPS-71-tab}.

Similarly we can determine top A-polynomials. We present the list of extremal A-polynomials for several torus knots in table \ref{tab-A-torus}. Note that for various knots these polynomials have the same terms of lower degree, and for higher degrees more terms appear for more complicated knots. It would be interesting to understand this pattern and its manifestation on the level of BPS numbers.

Furthermore, we observe that the Improved Integrality also holds for this family of torus knots, with the values of $\gamma^\pm$ given in table \ref{c-minmax-aug}. Note that the values of $\gamma^\pm$ grow linearly with $p$.


\begin{table}
\begin{small}
\be
\begin{array}{|c|cccccccc|} \hline
r \setminus i & 0 & 1 & 2 & 3 & 4 & 5 & 6 & 7  \\ \hline
1 & -2 & 3 & -1 & 0 & 0 & 0 & 0 & 0 \\
2 & 2 & -8 & 12 & -8 & 2 & 0 & 0 & 0 \\
3 & -3 & 27 & -84 & 126 & -99 & 39 & -6 & 0 \\
4 & 8 & -102 & 488 & -1214 & 1764 & -1554 & 816 & -234 \\
5 & -26 & 413 & -2682 & 9559 & -20969 & 29871 & -28203 & 17537 \\
6 & 90 & -1752 & 14484 & -67788 & 201810 & -405888 & 569322 & -564192 \\
7 & -329 & 7686 & -77473 & 451308 & -1711497 & 4499696 & -8504476 & 11792571 \\
8 & 1272 & -34584 & 411948 & -2882152 & 13350352 & -43658370 & 104759240 & -188904738 \\
9 & -5130 & 158730 & -2183805 & 17877558 & -98157150 & 385713186 & -1128850632 & 2524827921 \\
10 & 21312 & -740220 & 11560150 & -108550256 & 690760044 & -3179915704 & 11028120884 & -29597042376 \\ \hline
\end{array}
\nonumber
\ee
\caption{BPS invariants $b_{r,i}$ for the $3_1$ knot.} 
\label{aBPS-31-tab}
\end{small}
\end{table}

\begin{table}
\begin{small}
\be
\begin{array}{|c|cccc|} \hline
r \setminus i & 0 & 1 & 2 & 3  \\ \hline
1 & -3 & 5 & -2 & 0 \\
2 & 10 & -40 & 60 & -40 \\
3 & -66 & 451 & -1235 & 1750 \\
4 & 628 & -5890 & 23440 & -51978 \\
5 & -7040 & 83725 & -438045 & 1330465 \\
6 & 87066 & -1257460 & 8165806 & -31571080 \\
7 & -1154696 & 19630040 & -152346325 & 716238720 \\
8 & 16124704 & -315349528 & 2847909900 & -15779484560 \\
9 & -234198693 & 5179144365 & -53361940365 & 340607862518 \\
10 & 3508592570 & -86566211200 & 1002184712130 & -7243117544640 \\ \hline
\end{array}
\nonumber
\ee
\caption{BPS invariants $b_{r,i}$ for the $5_1$ knot.} 
\label{aBPS-51-tab}
\end{small}
\end{table}

\begin{table}
\begin{small}
\be
\begin{array}{|c|cccccccc|} \hline
r \setminus i & 0 & 1 & 2 & 3  & 4 & 5 & 6 & 7 \\ \hline
1 & -4 & 7 & -3 & 0 & 0 & 0 & 0 & 0 \\
2 & 28 & -112 & 168 & -112 & 28 & 0 & 0 & 0 \\
3 & -406 & 2618 & -6916 & 9604 & -7406 & 3010 & -504 & 0 \\
4 & 8168 & -71588 & 270928 & -579124 & 765576 & -641452 & 332864 & -97852 \\
5 & -193170 & 2139333 & -10554173 & 30562838 & -57563814 & 73721676 & -65048368 & 39063778 \\ \hline
\end{array}
\nonumber
\ee
\caption{BPS invariants $b_{r,i}$ for the $7_1$ knot.} 
\label{aBPS-71-tab}
\end{small}
\end{table}


\subsection{BPS invariants from augmentation polynomials: $6_2$, $6_3$, $7_3$, $7_5$, $8_{19}$, $8_{20}$, $8_{21}$, $10_{124}$, $10_{132}$, and $10_{139}$ knots}   \label{ssec-aug}

Using the methods presented above we can easily compute BPS invariants for knots with known augmentation polynomials. Moreover, in many nontrivial cases we can confirm the conjecture that augmentation polynomials agree with Q-deformed polynomials (defined as the classical limit of recursion relations satisfied by colored HOMFLY polynomials) and $t=-1$ limit of super-A-polynomials. This conjecture has been explicitly verified for several torus and twist knots in \cite{AVqdef,superA,FGSS}, where it was shown that appropriate change of variables relates the two algebraic curves. For example, starting with the super-A-polynomial for figure-8 knot (\ref{Asuper41}), setting $t=-1$ and changing variables (note that it is not simply (\ref{MLnorm})) \cite{superA}
\be
Q = a,\qquad \beta = M, \qquad \alpha = L\frac{1 - \beta Q}{Q(1-\beta)},
\ee
we obtain (up to some irrelevant simple factor) the Q-deformed polynomial in the same form as in \cite{AVqdef}
\bea
A^{\textrm{Q-def}}(\alpha,\beta,Q) & = &   (\beta^2 - Q \beta^3) + (2 \beta - 2 Q^2 \beta^4 + Q^2 \beta^5 - 1) \alpha + \nonumber \\
& & + (1 - 2 Q \beta + 2 Q^2 \beta^4 - Q^3 \beta^5) \alpha^2 + Q^2 (\beta-1) \beta^2 \alpha^3 \,, \nonumber
\eea
where it was shown to match the augmentation polynomial.

\begin{figure}[ht]
\begin{center}
\includegraphics[scale=0.25]{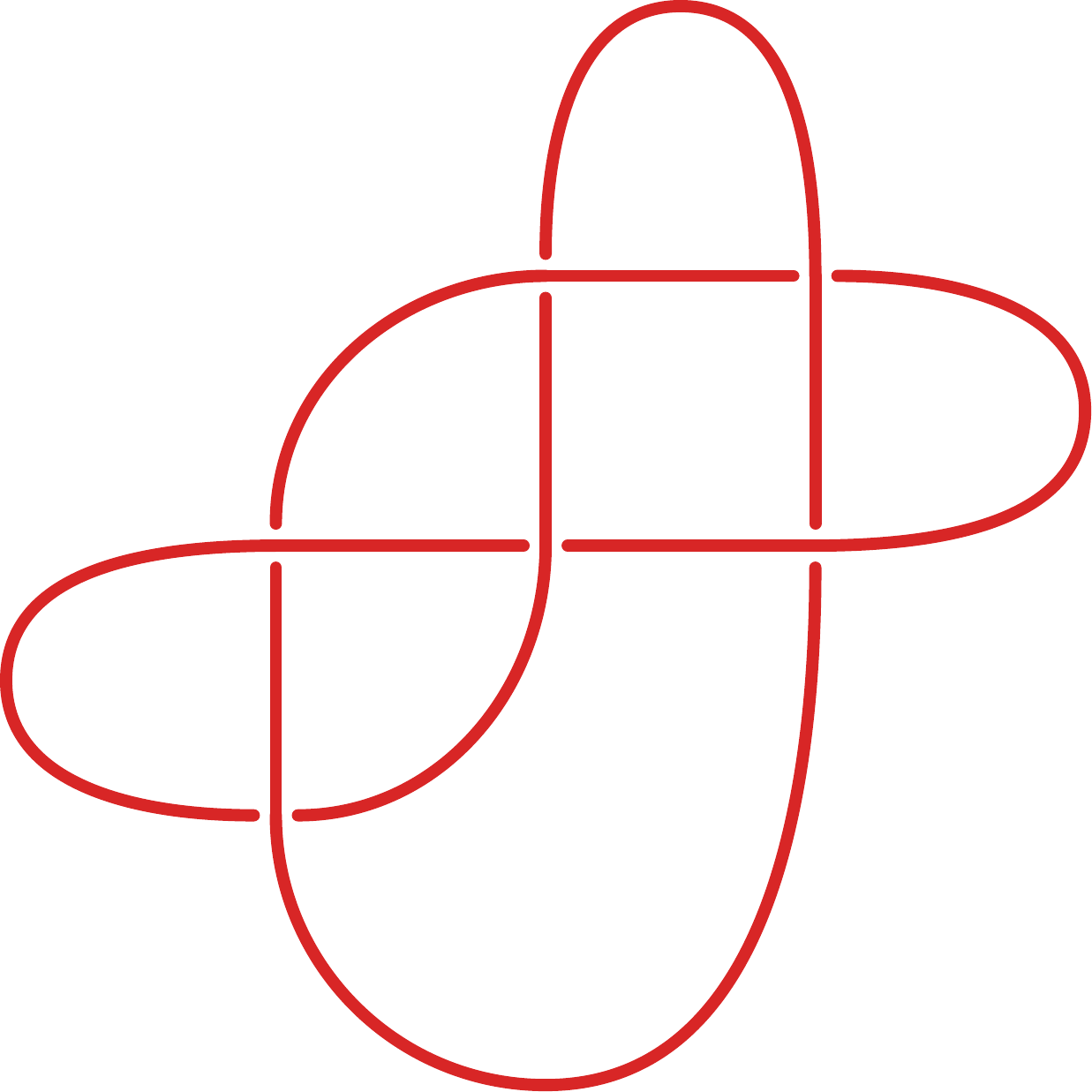}$\qquad$
\includegraphics[scale=0.25]{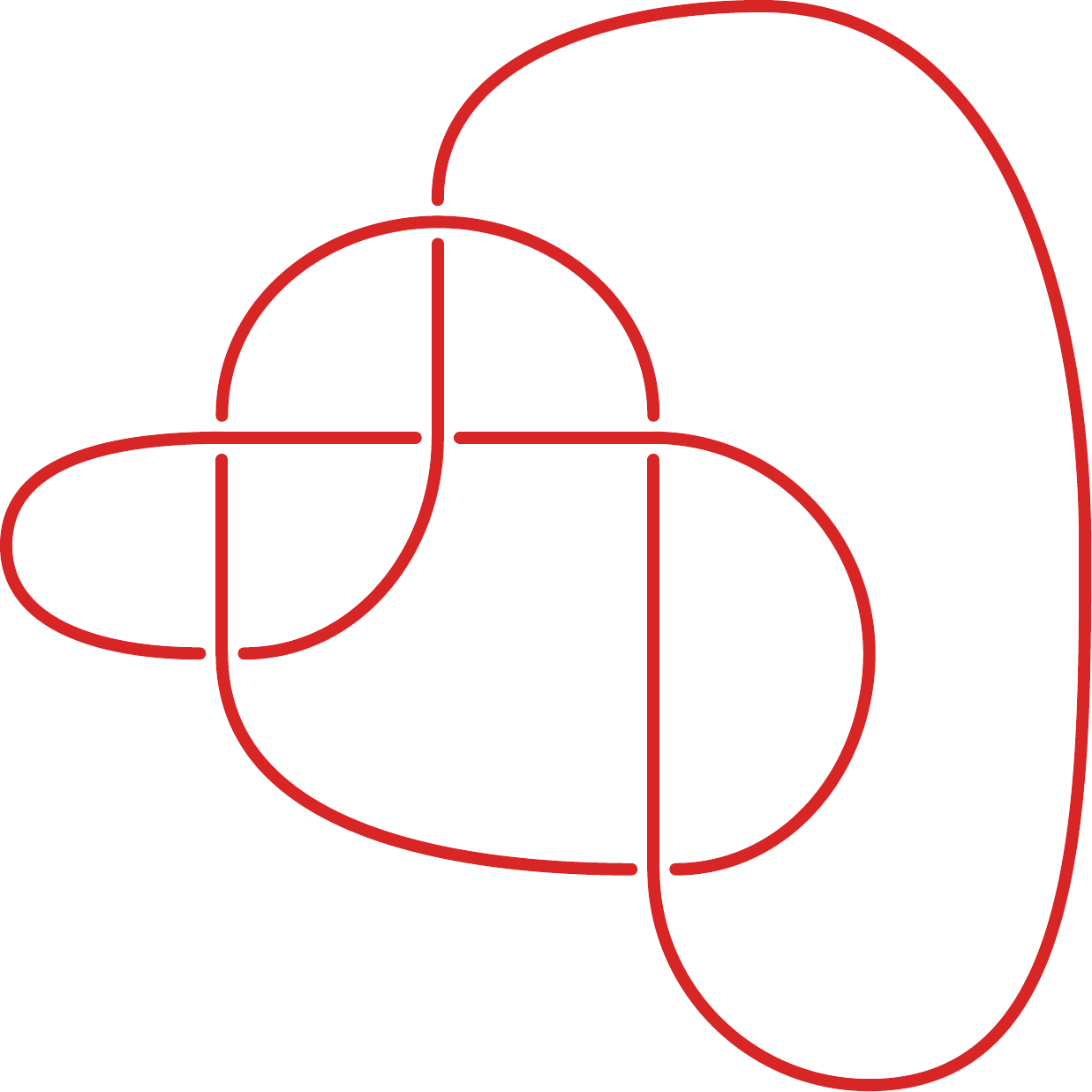}$\qquad$
\includegraphics[scale=0.25]{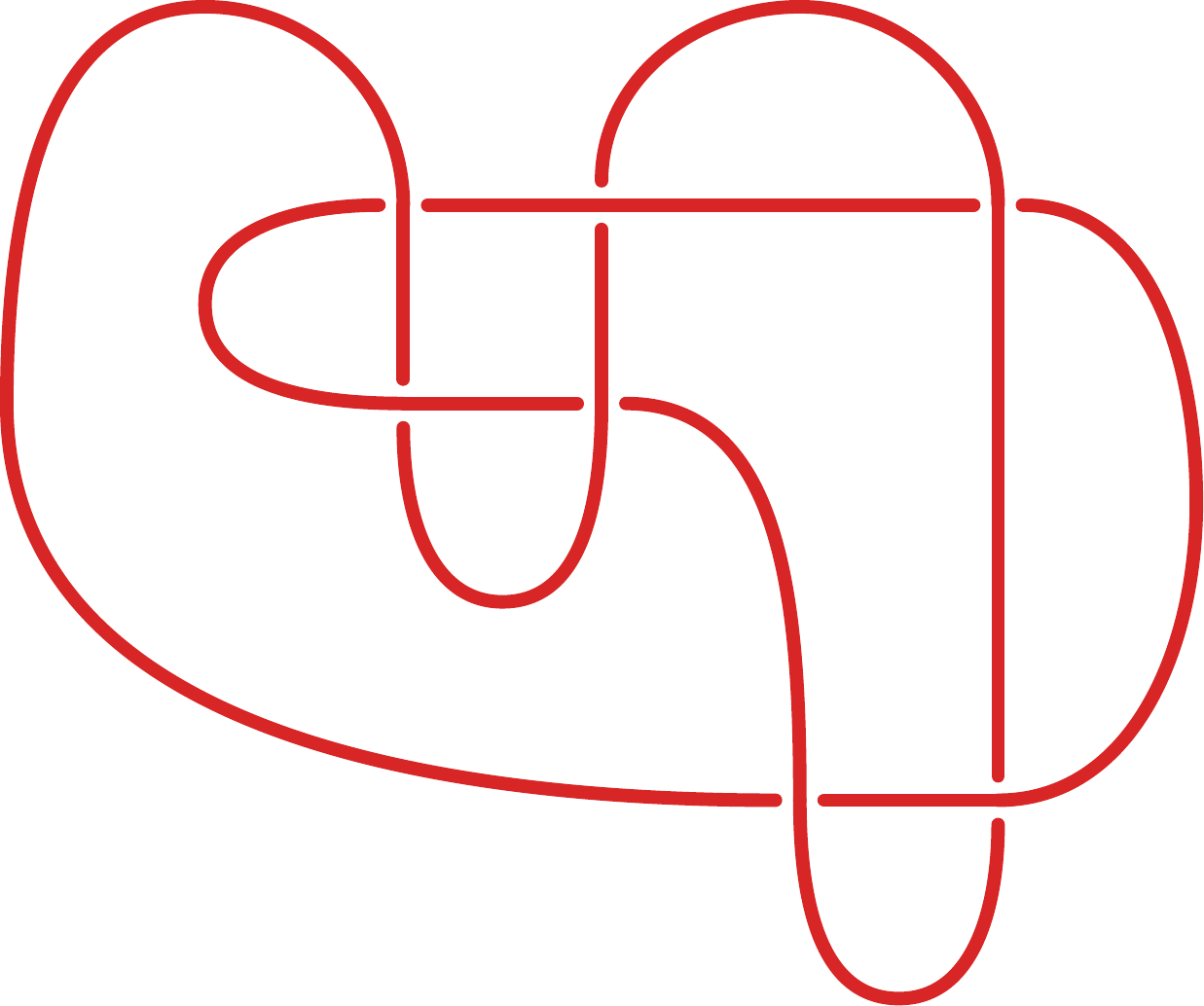}$\qquad$
\includegraphics[scale=0.25]{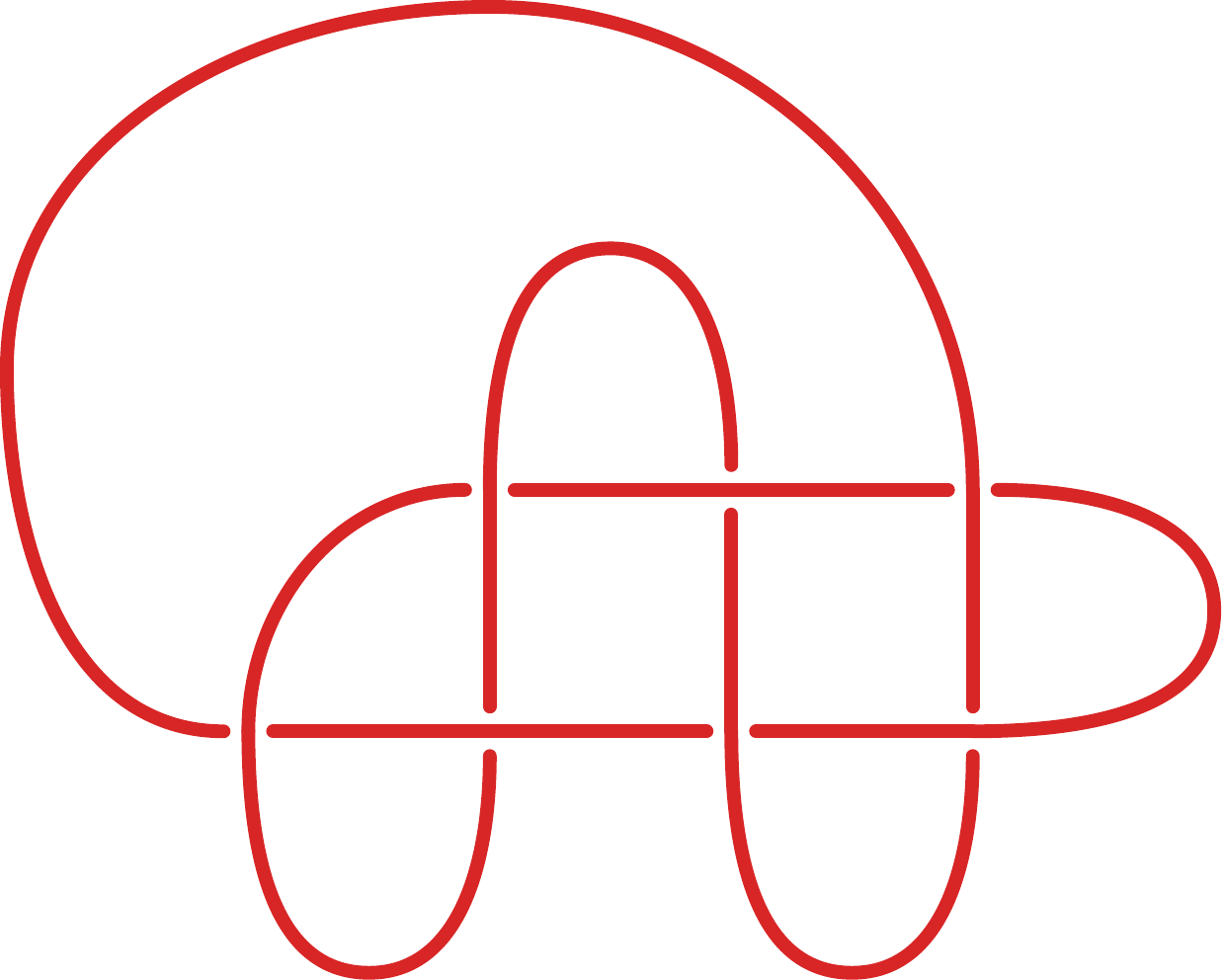}
\end{center}
\caption{The $6_2$, $6_3$, $7_3$ and $7_5$ knots.}
\label{fig-6-7-knots}
\end{figure}

Now we show that this conjecture can be verified for many non-trivial knots, even if an explicit form of Q-deformed polynomials or super-A-polynomial is not known. Let us consider the following knots: $6_2$, $6_3$, $7_3$, $7_5$, $8_{19}$, $8_{20}$, $8_{21}$, $10_{124}$, $10_{132}$, and $10_{139}$, for which augmentation polynomials are determined in \cite{Ng,NgFramed}. Changing the variables into $M$ and $L$ relevant for A-polynomials, and further into $x$ and $y$ relevant for our considerations, and from appropriate rescalings (\ref{Amin}) we obtain extremal A-polynomials. They are presented in table \ref{Aminmax-aug}, and constitute one of our main results.

From extremal polynomials in table \ref{Aminmax-aug} we can determine the extremal BPS invariants $b^{\pm}_r$ for arbitrary $r$; some results are presented in tables in appendix \ref{app-bps}. We also experimentally confirm the Improved Integrality -- the constants $\gamma^\pm$ for some knots are shown in table \ref{c-minmax-aug}. However, unfortunately, the corresponding functions $Y^\pm(x)$ no longer satisfy the 
fortunate condition \eqref{eq.fortunate}.

\begin{figure}[ht]
\centering
\includegraphics[scale=0.3]{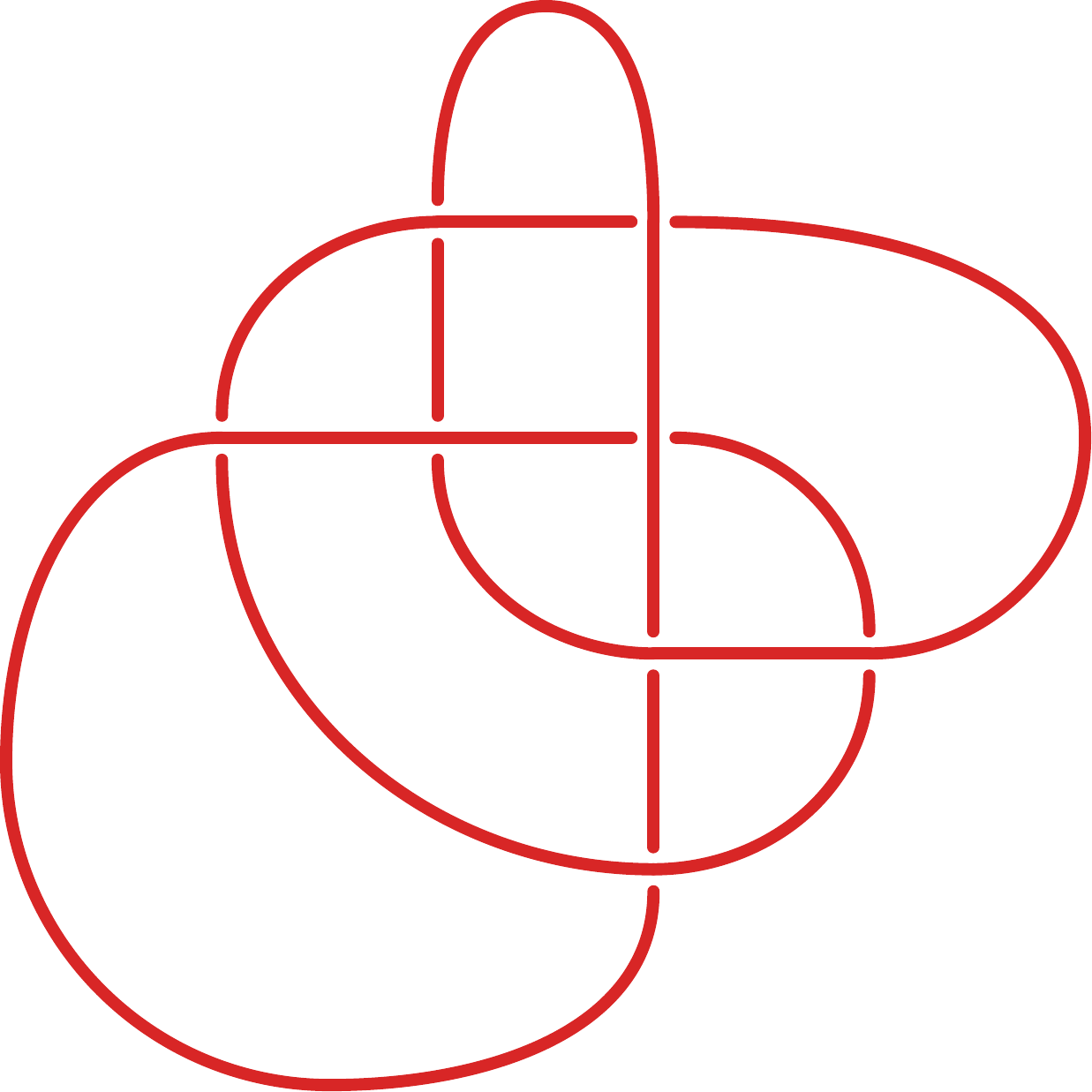}$\qquad$
\includegraphics[scale=0.3]{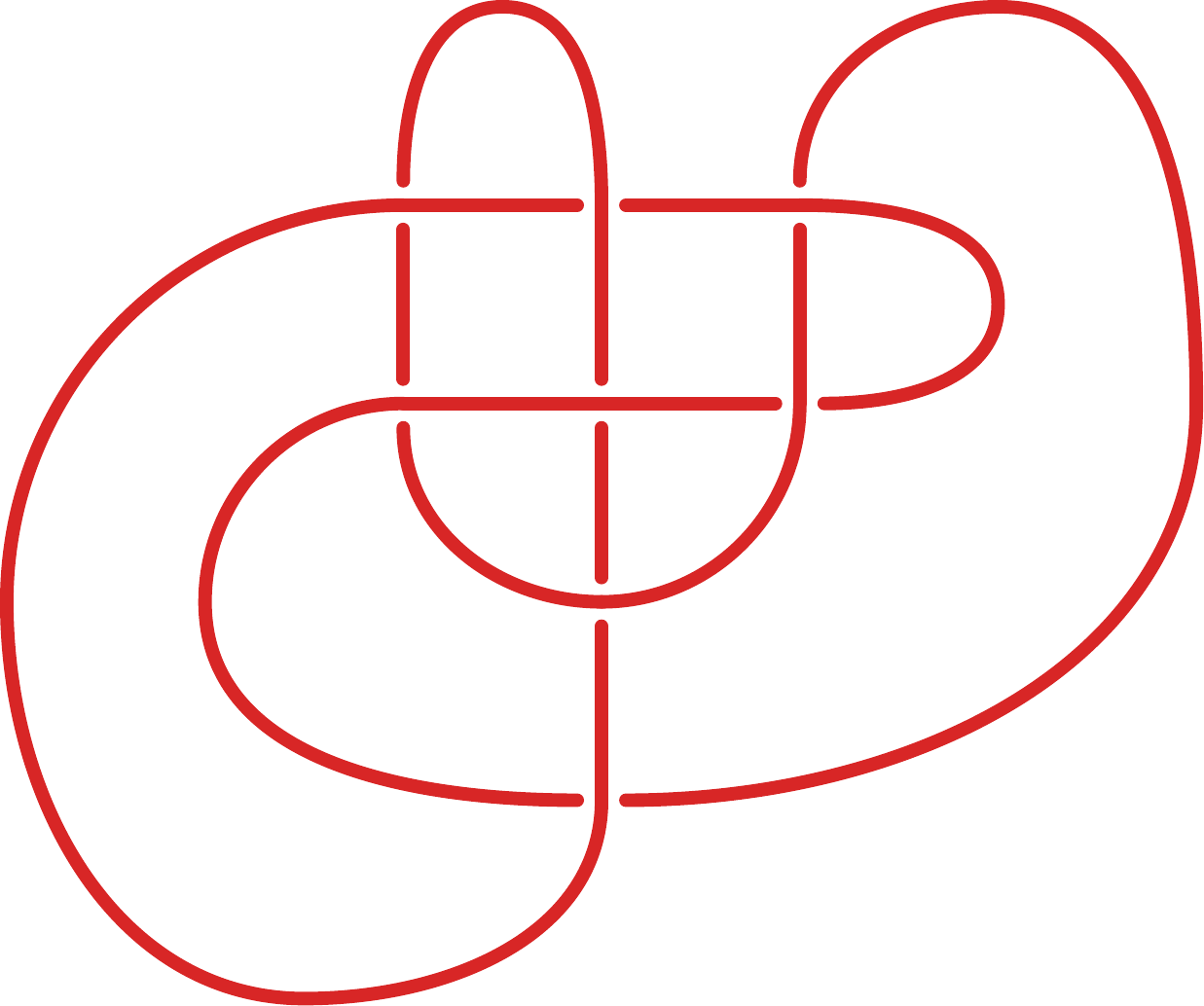}$\qquad$
\includegraphics[scale=0.3]{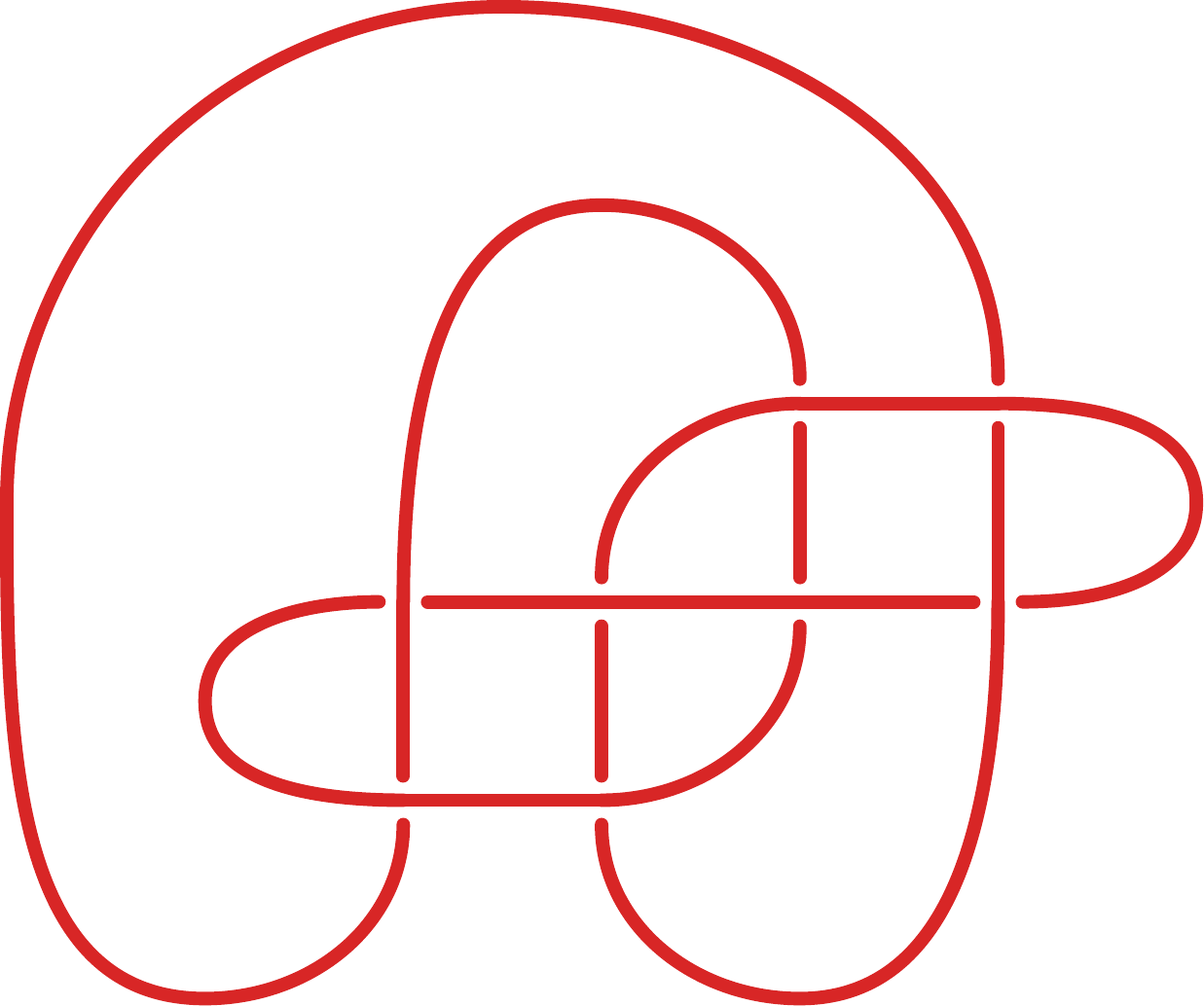}
\caption{The $8_{19}$, $8_{20}$ and $8_{21}$ knots.}
\label{fig-8-knots}
\end{figure}

Even though the Q-deformed or super-A-polynomials are not known for knots listed in table \ref{Aminmax-aug}, colored HOMFLY polynomials for those knots for several values of $r$ have been explicitly determined in \cite{Nawata:2013qpa,Wedrich:2014zua}. We can therefore determine the corresponding invariants $N_{r,i,j}$ using LMOV formulas (\ref{f-P}) -- we list some of these invariants in tables in appendix \ref{app-lmov}. On the other hand, from the known augmentation polynomials we can compute some BPS invariants $b_{r,i}$ using our techniques. In all cases we find the agreement between these two computations, as the reader can also verify by comparing tables in appendices \ref{app-bps} and \ref{app-lmov}. This is quite a nontrivial test of the (still conjectural) relation between augmentation polynomials and colored HOMFLY polynomials.

For example, consider the LMOV invariants $N_{r,i,j}$ of the $6_2$ knot for $r=1,2,3$, given in tables \ref{62-N1ij}, \ref{62-N2ij}, \ref{62-N3ij}. We determined these invariants from the knowledge of HOMFLY polynomials, determined up to $r=4$ in \cite{Nawata:2013qpa}, and applying formulas (\ref{f-P}). To obtain $b^-_r$ and $b^+_r$ we need to resum, respectively, the first and the last row in those tables (corresponding to minimal and maximal power of $a$). For the minimal case (first rows in the tables) from the resummation we obtain the number $-1,-2,-10$, and for the maximal case (last rows in the tables) we find the numbers $2,2,7$. These results indeed agree with values of $b^\pm_r$ for $6_2$ knot given in table \ref{6-7-br-aug}, which are determined from its augmentation polynomial (more precisely, the corresponding extremal A-polynomials given in table \ref{Aminmax-aug}). We verified such an agreement for other knots discussed in this section.

\begin{figure}[ht]
\centering
\includegraphics[scale=0.3]{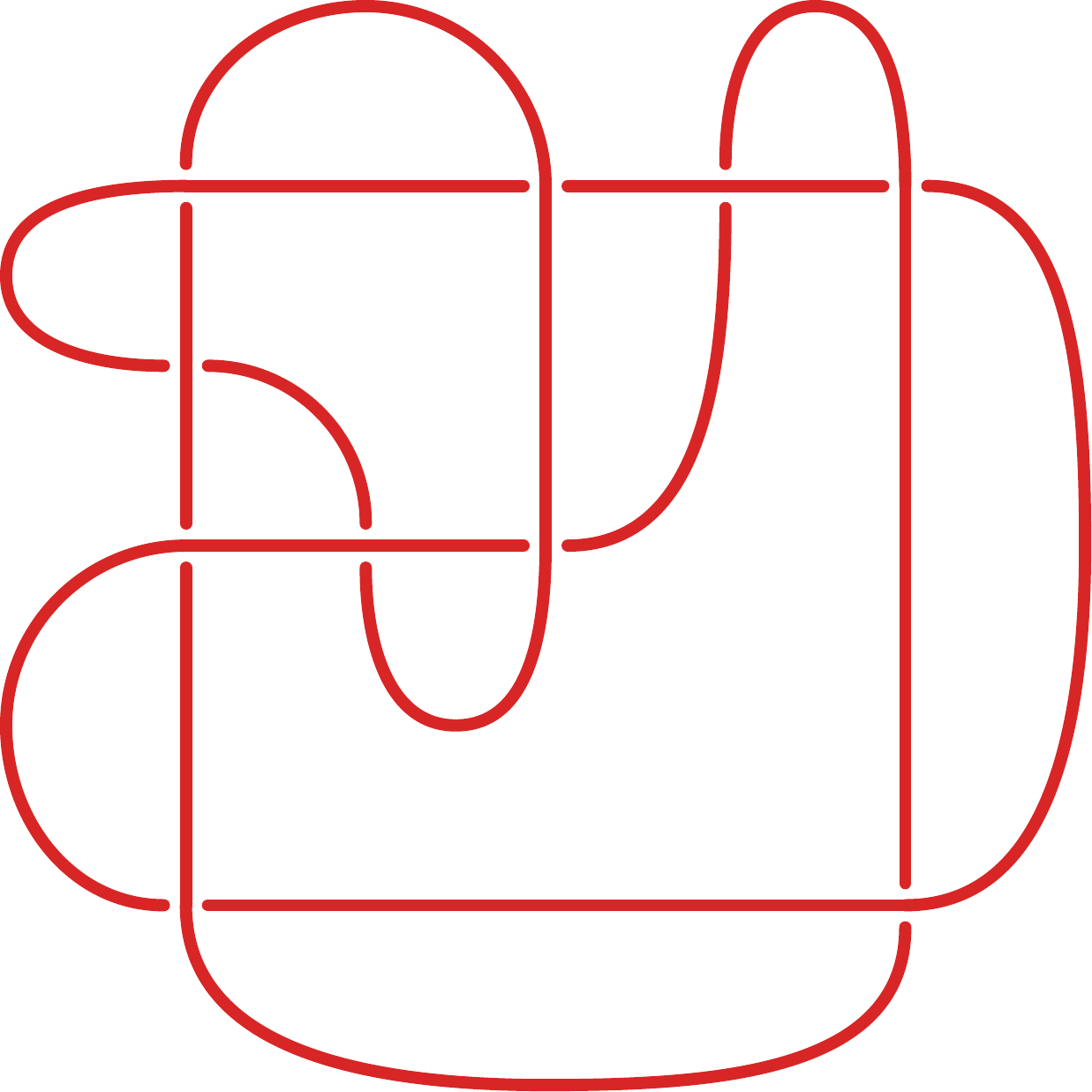}$\qquad$
\includegraphics[scale=0.3]{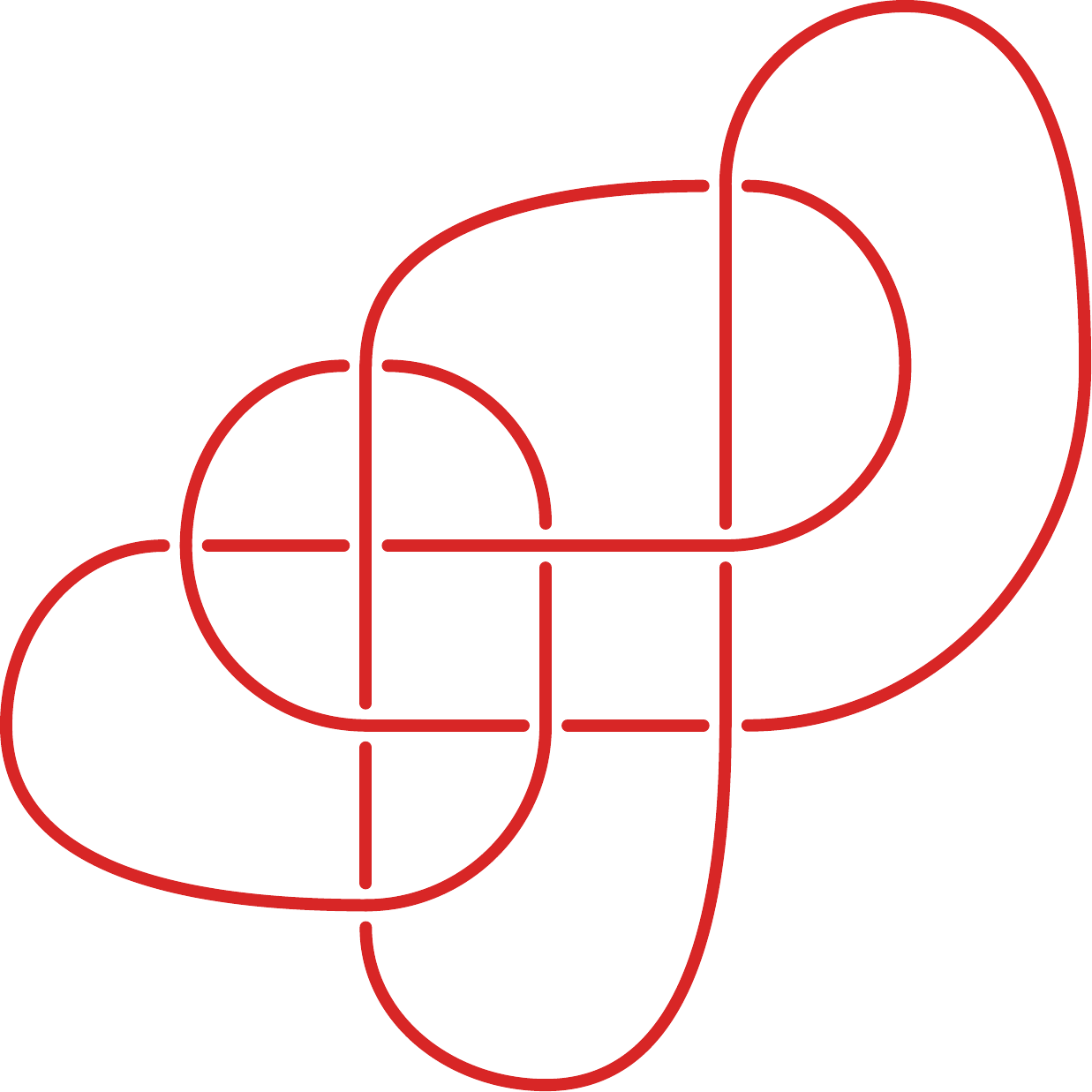}$\qquad$
\includegraphics[scale=0.3]{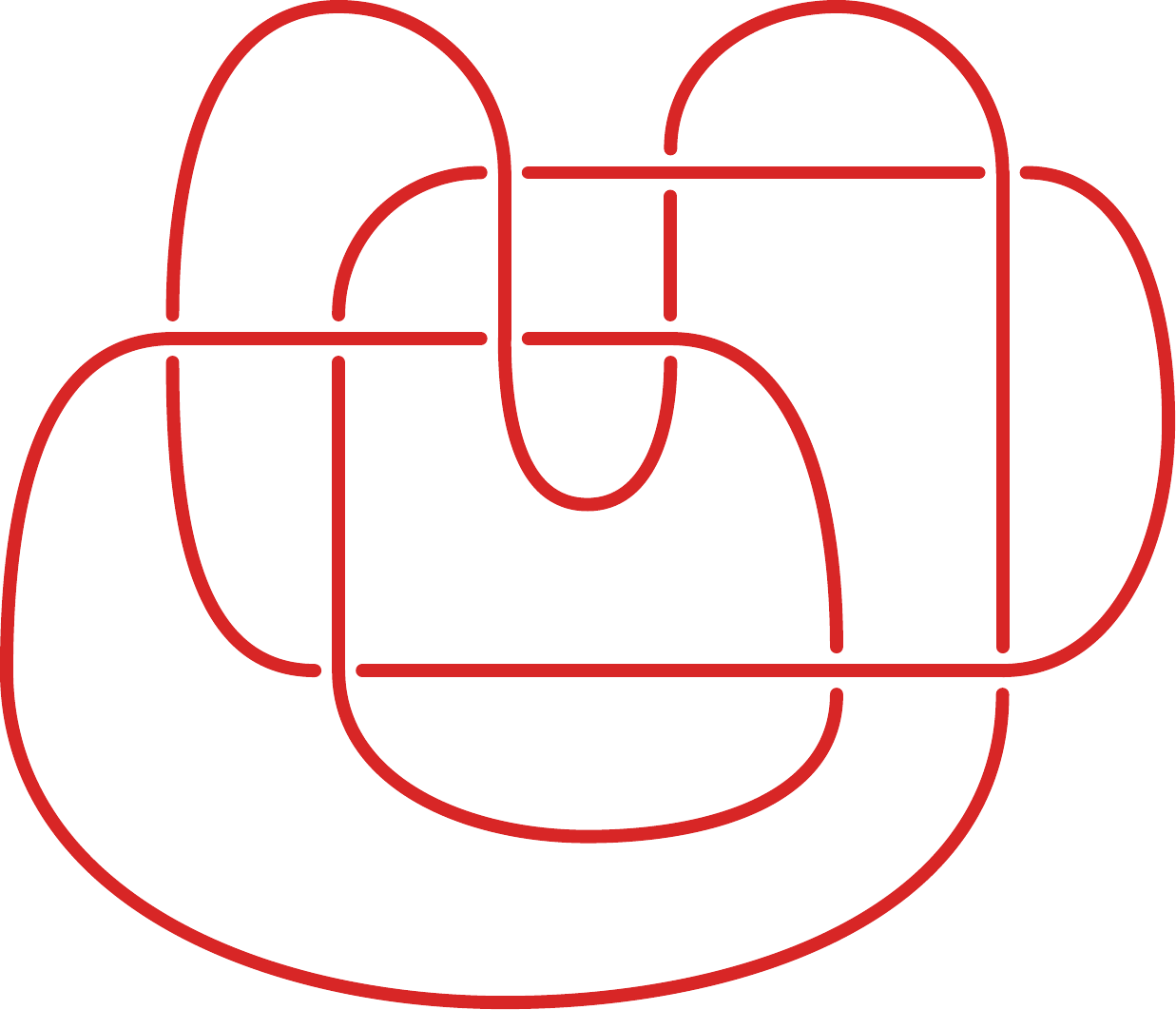}
\caption{The $10_{124}$, $10_{132}$ and $10_{139}$ knots.}
\label{fig-10-knots}
\end{figure}

%
%
%
%


\subsection{Refined BPS invariants from super-A-polynomials}

Beyond the $a$-dependence we can consider further deformation of A-polynomials, in parameter $t$, that leads to super-A-polynomials. It is natural to ask if super-A-polynomials encode refined BPS degeneracies $b_{r,i,j}$. Such degeneracies should be identified with generalized LMOV invariants, defined by relations (\ref{Pr-LMOVref}), and could be determined from the knowledge of super-A-polynomials, using the relation (\ref{xdyy-ref}). This conjecture would be confirmed if $b_{r,i,j}$ would turn out to be integer. In this section we extract such invariants from the known super-A-polynomials. To this end it is convenient to consider super-A-polynomials as $T$-deformation and $a$-deformation of bottom A-polynomials (that arise for $a=0$ and $T=1$), where $t=-T^2$. This results in slightly redefined degeneracies $\tilde{b}_{r,i,j}$, which can be combined into the generating functions
\be
\sum_{r,i,j} \tilde{b}_{r,i,j} x^r a^i T^j.
\ee
We present such generating functions in table \ref{tab-refBPS}. Clearly all coefficients in these generating functions are integer, and therefore capture putative refined BPS degeneracies. We plan to analyze these refined BPS invariants for knots in more detail in future work.

\begin{table}
\begin{small}
\be
\begin{array}{|c|l|}
\hline 
\textrm{\bf Knot} & \qquad \qquad\qquad  \sum_{r,i,j} \tilde{b}_{r,i,j} x^r a^i T^j   \nonumber \\
\hline 
\hline
{\bf 4_1} & (-1 + a T - a T^2 + 2 a T^3 - 2 a^2 T^4 + a^2 T^5) x + (-1 + 2 a T + (-a - a^2) T^2 + (3 a + a^2) T^3 + \\
              & -  4 a^2 T^4 + (a^2 + a^3) T^5) x^2 + (-3 + 7 a T + (-3 a - 5 a^2) T^2 + (10 a + 5 a^2 + a^3) T^3 + \\
              & + (-21 a^2 - 2 a^3) T^4 + (6 a^2 +  13 a^3) T^5) x^3 + (-10 + 30 a T + (-12 a - 32 a^2) T^2 + \\
             & + (42 a + 28 a^2 + 14 a^3) T^3 + (-117 a^2 - 21 a^3 - 2 a^4) T^4 + (35 a^2 + 114 a^3 + 5 a^4) T^5) x^4 +\\
             & + (-40 + 143 a T + (-55 a - 198 a^2) T^2 + (198 a + 165 a^2 + 132 a^3) T^3 + \\
             & + (-690 a^2 - 180 a^3 - 42 a^4) T^4 + (210 a^2 + 912 a^3 + 84 a^4 + 5 a^5) T^5) x^5 +\dots \\
\hline
{\bf 6_1} &  (-1 + a T - 2 a T^2) x + (-1 + 2 a T + (-3 a - a^2) T^2) x^2 + (-3 + 7 a T + (-10 a - 5 a^2) T^2) x^3 +\\ 
             & + (-10 + 30 a T + (-42 a - 32 a^2) T^2) x^4 + (-40 + 143 a T + (-198 a - 198 a^2) T^2) x^5 + \dots \\
\hline
{\bf 5_2} & (-1 + T + (-1 - a) T^2 + 2 a T^3 - 2 a T^4 + (a + 2 a^2) T^5) x + (T^2 + (-1 - 2 a) T^3 + \\
              & + (1 + 5 a + a^2) T^4 + (-8 a - 4 a^2) T^5) x^2 + (T^3 + (-3 - 4 a) T^4 + (2 + 16 a + 4 a^2) T^5) x^3 + \\
              & + (2 T^4 + (-6 - 10 a) T^5) x^4 + 4 T^5 x^5 + \dots \\
\hline
{\bf 3_1} & (-1 - T^2 + 2 a T^3 + a T^5) x + (T^2 - a T^3 + T^4 - 5 a T^5) x^2 +\\
              & + (-2 T^4 + 7 a T^5) x^3 + (T^4 - 3 a T^5) x^4 + \dots \\
\hline
{\bf 5_1} & (-1 - T^2 + 2 a T^3 - T^4) x + (T^2 - a T^3 + 4 T^4) x^2 - 5 T^4 x^3 + 2 T^4 x^4 +\dots \\
\hline  
\end{array}
\ee  
\end{small}
\caption{Generating functions  of refined degeneracies $\tilde{b}_{r,i,j}$ for several knots. Integrality of coefficients confirms the refined version of the LMOV conjecture.} \label{tab-refBPS}
\end{table}


\section{Conclusions and discussion}  \label{sec-conclude}

The results of this work deserve further studies that we plan to undertake. On one hand they should inspire mathematical research. We have formulated and tested various conjectures, in particular divisibility by $r^2$ following from the conjectured LMOV integrality, and Improved Integrality of extremal BPS degeneracies.  These statements should hold for all knots and proving them is an important task (even proofs of divisibility by $r^2$ in various specific cases, in particular (\ref{br-neg-intro}) and (\ref{br-pos-intro}), are challenging); note that proofs of integrality of Gopakumar-Vafa in certain cases were given in \cite{Kontsevich:2006an,Vologodsky:2007ef,Schwarz:2008ti}, and presumably these techniques could be generalized to the case of knots. We also associated new integer invariants $\gamma^{\pm}$, related to Improved Integrality, to all knots -- it is important to understand deeper their mathematical meaning and, possibly, relation to other characteristics of knots. Furthermore, for some knots we observed that the solutions of extremal A-polynomial equations are given by hypergeometric functions. It is important to understand for which knots such algebraic hypergeometric functions arise and if they have any further meaning. Understanding which knots have this property could lead to a new method to determine many other algebraic hypergeometric functions (associated to various knots). 

It would also be interesting to understand our results from the perspective of knot homologies. Currently the most powerful method to determine -- at least conjecturally -- colored HOMFLY homologies is the formalism of (colored) differentials \cite{DGR,GS}. These differentials reveal intricate structure not only of colored homologies, but also of ordinary HOMFLY invariants. Therefore they should capture some essential information about unrefined and refined BPS degeneracies that we consider. Note that the ``bottom row'' structure of HOMFLY homologies (i.e. corresponding to the minimal power of $a$) was analyzed e.g. in \cite{Gorsky:2013jxa}; it would be interesting to relate it to BPS degeneracies determined here.

Our results also raise further interesting questions on the physics side. First, in the introduction we already mentioned their intimate connection to 3d-3d duality, which relates knot invariants to 3-dimensional $\mathcal{N}=2$ theories \cite{DGG,superA,Chung:2014qpa}. Various objects in our analysis, such as colored knot polynomials, super-A-polynomials, etc., play an important role in this duality. Therefore all the new objects and statements that we consider should also find its interpretation on the $\mathcal{N}=2$ side of this duality. 

Second, the results such as Improved Integrality or formulation of refined BPS invariants for knots generalize the statements of the original LMOV conjectures \cite{OoguriV,Labastida:2000zp,Labastida:2000yw}. It is desirable to understand in more detail M-theory interpretation of these results. In particular we obtain the BPS degeneracies in an analogous way as has been done for D-branes in \cite{AV-discs,AKV-framing}. Furthermore, it has been conjectured in \cite{OoguriV,AVqdef} that all knots should be mirror to Lagrangian branes in the conifold geometry, and for some knots such Lagrangian branes have been constructed \cite{Diaconescu:2011xr,Jockers:2012pz}. It would be amusing to construct such Lagrangian branes for other knots that we consider, and compare the degeneracies they encode with our computations.

Third, our results concern primarily the classical algebraic curves, i.e. the $q\to 1$ limit of recursion relations for knot polynomials.  It is desirable to introduce the dependence on the parameter $q$ and determine corresponding BPS degeneracies directly from the knowledge of those recursion relations. Our results can also be further generalized to higher-dimensional varieties generalizing algebraic curves, and correspondingly to links or knots labeled by more general (multi-row) representations.

Fourth, an important challenge is to understand refined open BPS states that we compute for several knots, based on the known super-A-polynomials. Recently various formulations of closed refined BPS states have been considered, see e.g. \cite{Choi:2012jz,Huang:2013yta}. It would be nice to make contact between these various approaches involving open and closed BPS states.

Yet another intriguing direction of research relating algebraic curves and knot invariants has to do with the topological recursion. It has been conjectured in \cite{DijkgraafFuji-1} and further analyzed in \cite{DijkgraafFuji-2,Borot:2012cw,abmodel,BEM,Gu:2014yba} that the asymptotic expansion of colored Jones or HOMFLY polynomials can be reconstructed from the topological recursion for the A-polynomial curve. This conjecture have been tested in a very limited number of cases and it still seems poorly understood. The new algebraic curves that we consider in this paper, in particular extremal A-polynomials, should provide a simpler setup in which this conjecture can be analyzed. 



\acknowledgments{We thank Estelle Basor, Brian Conrey, Sergei Gukov, Maxim Kontsevich, Satoshi Nawata, and Marko Sto$\check{\text{s}}$i$\acute{\text{c}}$ for insightful discussions. We greatly appreciate hospitality of American Institute of Mathematics, Banff International Research Station, International Institute of Physics in Natal, and Simons Center for Geometry and Physics, where parts of this work were done. This work is supported by the ERC Starting Grant no. 335739 \emph{``Quantum fields and knot homologies''} funded by the European Research Council under the European Union's Seventh Framework Programme, and the Foundation for Polish Science.}


\newpage

\appendix

\section{Extremal $A$-polynomials for various knots}
\label{sec-extremeA}

\begin{table}[H]
\begin{small}
\be
\begin{array}{|c|c|c|}
\hline 
\textrm{\bf Knot} & \mathcal{A}^-(x,y) & \mathcal{A}^+(x,y)   \nonumber \\
\hline 
\hline
\ {\bf 0_1}=K_0 \ & 1-x-y^2 & 1+xy^2-y^2 \\
\hline
\ K_p,\ p\leq -1 \ & \quad x-y^4+y^6  \quad &  \quad 1-y^2+x y^{4|p|+2} \quad  \\
\hline
\ K_p,\ p\geq 2\  & \quad 1-y^2- x y^4  \quad & \quad   1-y^2+x y^{4p+4}  \quad  \\
\hline
\end{array}
\ee  
\end{small}
\caption{Extremal A-polynomials for twist knots $K_p$ (including the unknot $0_1= K_0$).} \label{tab-A-twists}
\end{table}

\begin{table}[H]
\begin{small}
\be
\begin{array}{|c|c|c|}
\hline 
\textrm{\bf Knot} & \mathcal{A}^-(x,y) & \mathcal{A}^+(x,y)   \nonumber \\
\hline 
\hline
{\bf 3_1} & -1 + x + y^2 - x y^2 + 2 x y^4 + x^2 y^6 & -1 + y^2 + x y^8 \\
\hline
{\bf 5_1} & -1 + x + y^2 - x y^2 + 2 x y^4 - 2 x y^6 +   &    -1 + y^2 + x y^8 - x y^{10} + 2 x y^{12} + x^2 y^{22} \\
&  + 3 x y^8 + 2 x^2 y^{10} - x^2 y^{12} + 3 x^2 y^{14} + x^3 y^{20} & \\
\hline
{\bf 7_1} & -1 + x + y^2 - x y^2 + 2 x y^4 - 2 x y^6 + 
 & -1 + y^2 + x y^8 - x y^{10} + 2 x y^{12} - 2 x y^{14} +  \\
&  + 3 x y^8 - 3 x y^{10} + 4 x y^{12} + 3 x^2 y^{14} - 2 x^2 y^{16}      &  +3 x y^{16} + 2 x^2 y^{26} - x^2 y^{28} + 3 x^2 y^{30} + x^3 y^{44}   \\
& + 6 x^2 y^{18} - 3 x^2 y^{20} + 6 x^2 y^{22} +  3 x^3 y^{28} +    &  \\
& - x^3 y^{30} + 4 x^3 y^{32} + x^4 y^{42} & \\
\hline
{\bf 9_1} &  -1 + x + y^2 - x y^2 + 2 x y^4 - 2 x y^6 +    &  -1 + y^2 + x y^8 - x y^{10} + 2 x y^{12} - 2 x y^{14}  \\
& + 3 x y^8 - 3 x y^{10} + 4 x y^{12} - 4 x y^{14} + 5 x y^{16} +    &   +3 x y^{16} - 3 x y^{18} + 4 x y^{20} + 3 x^2 y^{30}+  \\
& + 4 x^2 y^{18} - 3 x^2 y^{20} + 9 x^2 y^{22} - 6 x^2 y^{24} +   &-2 x^2 y^{32} + 6 x^2 y^{34} - 3 x^2 y^{36} + 6 x^2 y^{38}  +    \\
& + 12 x^2 y^{26}  -6 x^2 y^{28} + 10 x^2 y^{30} + 6 x^3 y^{36} +  &  +3 x^3 y^{52} - x^3 y^{54} + 4 x^3 y^{56} + x^4 y^{74} \\
& - 3 x^3 y^{38} + 12 x^3 y^{40} -4 x^3 y^{42} + 10 x^3 y^{44} + &      \\
& + 4 x^4 y^{54} - x^4 y^{56} + 5 x^4 y^{58} +  x^5 y^{72} & \\
\hline
{\bf 11_1}   & -1 + x + y^2 - x y^2 + 2 x y^4 - 2 x y^6 +   & -1 + y^2 + x y^8 - x y^{10} + 2 x y^{12} - 2 x y^{14} + \\
&+ 3 x y^8 - 3 x y^{10} + 4 x y^{12} - 4 x y^{14} +   & +3 x y^{16} - 3 x y^{18} + 4 x y^{20} - 4 x y^{22} +\\
&+ 5 x y^{16} - 5 x y^{18} + 6 x y^{20} + 5 x^2 y^{22} +   & +5 x y^{24} + 4 x^2 y^{34} - 3 x^2 y^{36} + 9 x^2 y^{38} + \\
&-4 x^2 y^{24}+ 12 x^2 y^{26} - 9 x^2 y^{28} + 18 x^2 y^{30} +   &-6 x^2 y^{40} + 12 x^2 y^{42} - 6 x^2 y^{44} + 10 x^2 y^{46} +   \\
&-12 x^2 y^{32} + 20 x^2 y^{34} - 10 x^2 y^{36} + 15 x^2 y^{38} +   & +6 x^3 y^{60} - 3 x^3 y^{62} + 12 x^3 y^{64} - 4 x^3 y^{66} + \\
&+10 x^3 y^{44} - 6 x^3 y^{46} + 24 x^3 y^{48} - 12 x^3 y^{50} +   & +10 x^3 y^{68} + 4 x^4 y^{86} - x^4 y^{88} + 5 x^4 y^{90} + x^5 y^{112} \\
&+30 x^3 y^{52} - 10 x^3 y^{54} + 20 x^3 y^{56} + 10 x^4 y^{66} +   & \\
&-4 x^4 y^{68} + 20 x^4 y^{70} - 5 x^4 y^{72} + 15 x^4 y^{74} +   & \\
&+5 x^5 y^{88} - x^5 y^{90} + 6 x^5 y^{92} + x^6 y^{110}   & \\
\hline  
\end{array}
\ee  
\end{small}
\caption{Extremal A-polynomials for torus knots.} \label{tab-A-torus}
\end{table}

\newpage

\begin{table}[H]
\begin{small}
\be
\begin{array}{|c|c|c|}
\hline 
\textrm{\bf Knot} & \mathcal{A}^-(x,y) & \mathcal{A}^+(x,y)   \nonumber \\
\hline 
\hline
{\bf 6_2} & x - y^8 + y^{10} & 1 - y^2 + 2 x y^2 - x y^4 + x^2 y^4 + x y^6 \\
\hline
{\bf 6_3} &  x + y^2 - y^4 &  -1 + y^2 + x y^4 \\
\hline
{\bf 7_3} & -1 + y^2 + x y^8 & -1 + y^2 + x y^{12} - x y^{14} + 2 x y^{16} + x^2 y^{30} \\
\hline
{\bf 7_5} & x + y^{14} - y^{16} & x^2 + 2 x y^6 - x y^8 + x y^{10} + y^{12} - y^{14} \\
\hline
{\bf 8_{19} } &  -1 + x + y^2 - x y^2 + 3 x y^6 + x y^8  - 4 x y^{10} +      & -1 + y^2 - x y^{18} \\
&  + 5 x y^{12} + x^2 y^{12} -  x^2 y^{14} + 5 x^2 y^{16} +3 x^2 y^{18} +  & \\
&   - 6 x^2 y^{20} + 10 x^2 y^{22} -  x^3 y^{24} + x^3 y^{26} +& \\
& + 3 x^3 y^{28} - 4 x^3 y^{30} + 10 x^3 y^{32} -  x^4 y^{36}  +&  \\
& + x^4 y^{38} - x^4 y^{40} + 5 x^4 y^{42} + x^5 y^{52} & \\
\hline
{\bf 8_{20} } &  x^2 - x y^4 + x y^6 - 2 x y^8 - y^{14} + y^{16} &  -x + x^2 - y^2 + 2 x y^2 - 5 x^2 y^2 + 4 x^3 y^2 + \\
& & + y^4 - x y^4 + 11 x^2 y^4 -  9 x^3 y^4 + 6 x^4 y^4 + \\
& & + x y^6 - 7 x^2 y^6 + 17 x^3 y^6 - 7 x^4 y^6 +  4 x^5 y^6 + \\
& & + 4 x^2 y^8 - 12 x^3 y^8 + 9 x^4 y^8 - 2 x^5 y^8 +  x^6 y^8 + \\
& & + 5 x^3 y^{10} - 6 x^4 y^{10} + x^5 y^{10} - x^3 y^{12} + 2 x^4 y^{12} -  x^5 y^{12} \\
\hline
{\bf 8_{21} } & x - y^8 + y^{10} &  x^4 + 3 x^3 y^2 + 3 x^2 y^4 - 7 x^3 y^4 + x y^6 - 10 x^2 y^6 + \\ 
& & +  2 x^3 y^6 - 3 x y^8 + 17 x^2 y^8 - x^3 y^8 - y^{10} + 2 x y^{10} + \\
& & - 11 x^2 y^{10} + y^{12} - 3 x y^{12} + 6 x^2 y^{12} - x^2 y^{14} \\
\hline
{\bf 10_{124} } & x^2 - 2 x y^{20} + x y^{22} - x y^{24} + y^{40} - y^{42} & x^7 + 7 x^6 y^{14} - x^6 y^{16} +  x^6 y^{18} - 2 x^6 y^{20} + \\
& & + 21 x^5 y^{28} -  6 x^5 y^{30} + 5 x^5 y^{32} - 3 x^5 y^{34} - x^5 y^{38} + \\
& & + x^5 y^{40} +  35 x^4 y^{42} - 15 x^4 y^{44} + 10 x^4 y^{46} + 8 x^4 y^{48} + \\
& & - 3 x^4 y^{50} -  3 x^4 y^{52} + 35 x^3 y^{56} - 20 x^3 y^{58} + 10 x^3 y^{60} + \\
& & + 22 x^3 y^{62} -  9 x^3 y^{64} + 2 x^3 y^{68} + 21 x^2 y^{70} - 2 x^3 y^{70} + \\
& & - 15 x^2 y^{72} +  5 x^2 y^{74} + 18 x^2 y^{76} - 9 x^2 y^{78} + 5 x^2 y^{80} + \\
& & +3 x^2 y^{82} +  7 x y^{84} - 6 x y^{86} + x y^{88} + 5 x y^{90} + \\
& & - 3 x y^{92} + 3 x y^{94} +  y^{98} - x y^{98} - y^{100} + x y^{100} \\
\hline
{\bf 10_{132} } & x^2 - x y^8 + x y^{10} - 2 x y^{12} - y^{22} + y^{24} & 1 - 4 x^2 + 6 x^4 - 4 x^6 + x^8 - y^2 + 6 x^2 y^2 - 12 x^4 y^2 + \\
& &  + 6 x^6 y^2 - 4 x^2 y^4 + 13 x^4 y^4  - 2 x^6 y^4 - 7 x^4 y^6 + x^4 y^8 \\
\hline
{\bf 10_{139} } & 1 - x - y^2 + x y^2 - 6 x y^6 + 10 x y^8 +& 1 - y^2 + x y^{22} \\
& - 9 x y^{10} + 2 x^2 y^{12} + 8 x y^{14} - 9 x^2 y^{14} - 9 x y^{16} + & \\
& + 17 x^2 y^{16} - 19 x^2 y^{18} -  16 x^2 y^{20} + 45 x^2 y^{22}  + & \\
& - 51 x^2 y^{24}- x^3 y^{24} + 8 x^3 y^{26} + 28 x^2 y^{28} +& \\
&  - 26 x^3 y^{28} - 36 x^2 y^{30} + 52 x^3 y^{30} -  61 x^3 y^{32} + & \\
& - 3 x^3 y^{34} + 75 x^3 y^{36} - 117 x^3 y^{38} - x^4 y^{38} + & \\
& 13 x^4 y^{40} + 56 x^3 y^{42} - 39 x^4 y^{42} - 84 x^3 y^{44} + & \\ 
&+ 65 x^4 y^{44} - 72 x^4 y^{46} + 28 x^4 y^{48} + 50 x^4 y^{50} + & \\
& + x^5 y^{50} -  135 x^4 y^{52} - 4 x^5 y^{52} + 12 x^5 y^{54} + & \\
& + 70 x^4 y^{56} -  23 x^5 y^{56} - 126 x^4 y^{58} + 37 x^5 y^{58} + & \\
& - 40 x^5 y^{60} +  32 x^5 y^{62} - 75 x^5 y^{66} + x^6 y^{68} + & \\
& + 56 x^5 y^{70} - 2 x^6 y^{70} -  126 x^5 y^{72} + 8 x^6 y^{72} + & \\
& - 13 x^6 y^{74} + 12 x^6 y^{76} -  15 x^6 y^{78} - 9 x^6 y^{80} +& \\
& + 28 x^6 y^{84} - 84 x^6 y^{86} - 3 x^7 y^{88} +  x^7 y^{90} + & \\
&- 5 x^7 y^{92} + 9 x^7 y^{94} + 8 x^7 y^{98} - 36 x^7 y^{100} + & \\
& 3 x^8 y^{108} + x^8 y^{112} - 9 x^8 y^{114} - x^9 y^{128} &  
\\
\hline  
\end{array}
\ee  
\end{small}
\caption{Extremal A-polynomials for various knots.} \label{Aminmax-aug}
\end{table}

\section{Extremal BPS invariants for various knots} 
\label{app-bps}



\begin{table}[H]
\be
\begin{array}{|c|c|c|c|c|c|c|c|c|}
\hline 
r & b^{-}_{6_2,r} & b^{+}_{6_2,r} & b^{-}_{6_3,r} & b^{+}_{6_3,r} & b^{-}_{7_3,r} & b^{+}_{7_3,r} & b^{-}_{7_5,r} & b^{+}_{7_5,r} \nonumber \\
\hline 
 1 & -1 & 2 & 1 & -1 & -1 & -2 & 1 & 2\\
 2 & -2 & 2 & -1 & 1 & 2 & 14 & -4 & -6\\
 3 & -10 & 7 & 1 & -1 & -6 & -183 & 28 & 31\\
 4 & -60 & 28 & -2 & 2 & 28 & 3\, 316 & -280 & -236\\
 5 & -425 & 134 & 5 & -5 & -155 & -70\, 502 & 3\, 290 & 2\, 118 \\
 6 & -3\, 296 & 695 & -13 & 13 & 936 & 1\, 656\, 429 & -42\, 616 & -20\, 923\\
 7 & -27\, 447 & 3\, 892 & 35 & -35 & -6\, 041 & -41\, 726\, 720 & 591\, 626 & 221\, 522\\
 8 & -240\, 312 & 22\, 888 & -100 & 100 & 41\, 080 & 1\, 106\, 293\, 848 & -8\, 644\, 784 & -2\, 469\, 560 \\
 9 & -2\, 188\, 056 & 140\, 139 & 300 & -300 & -290\, 565 & -30\, 500\, 237\, 331 & 131\, 347\, 227 & 28\, 632\, 747 \\
 \, 10\,  & \, -20\, 544\, 450 \, & \, 885\, 014 \, & \,  -925 & 925 \, & \, 2\, 119\, 190 \, & \, 867\, 235\, 597\, 322 \, & \, -2\, 058\, 115\, 960 \, & \, -342\, 395\, 810 \, \\
\hline  
\end{array}
\ee  
\caption{Extremal BPS invariants for the $6_2$, $6_3$, $7_3$ and $7_5$ knots. 
}      \label{6-7-br-aug}
\end{table}

\begin{table}[H]
\be
\begin{array}{|c|c|c|c|c|c|c|}
\hline 
r & b^{-}_{8_{19},r} & b^{+}_{8_{19},r} & b^{-}_{8_{20},r} & b^{+}_{8_{20},r} & b^{-}_{8_{21},r} & b^{+}_{8_{21},r} \nonumber \\
\hline 
 1 & -5 & 1 & 2 & -1 & -1 & 3 \\
 2 & 44 & 4 & -9 & -1 & -2 & -2 \\
 3 & -795 & 36 & 73 & 2 & -10 & 5 \\
 4 & 19\, 828 & 408 & -842 & 0 & -60 & -10 \\
 5 & -581\, 230 & 5\, 430 & 11\, 421 & -9 & -425 & 44 \\
 6 & 18\, 855\, 192 & 79\, 704 & -171\, 043 & 22 & -3\, 296 & -141 \\
 7 & -656\, 198\, 799 & 1\, 254\, 582 & 2\, 746\, 933 & 14 & -27\, 447 & 605 \\
 8 & 24\, 043\, 189\, 592 & 20\, 779\, 440 & -46\, 443\, 364 & -228 & -240\, 312 & -2\, 372 \\
 9 & -916\, 260\, 658\, 326 & 357\, 870\, 825 & 816\, 652\, 818 & 474 & -2\, 188\, 056 & 10\, 601 \\
 \, 10 \, & \, 36\, 017\, 403\, 983\, 220 \, & \, 6\, 356\, 271\, 400 \, & \, -14\, 811\, 255\, 141 \, & \, 1\, 008 \,  & \, -20\, 544\, 450 & -46\, 225 \, \\
\hline  
\end{array}
\ee  
\caption{Extremal BPS invariants for the $8_{19}$, $8_{20}$ and $8_{21}$ knots. 
}      \label{8 br_min_max_from_aug}
\end{table}

\begin{table}[H]
\be
\begin{array}{|c|c|c|c|c|c|c|}
\hline 
r & b^{-}_{10_{124},r} & b^{+}_{10_{124},r} & b^{-}_{10_{132},r} & b^{+}_{10_{132},r}& b^{-}_{10_{139},r} & b^{+}_{10_{139},r} \nonumber \\
\hline 
 1 & 2 & 7 & 2 & 0 & -6 & 1 \\
 2 & -13 & -119 & -13 & -1 & 87 & 5 \\
 3 & 157 & 4\, 194 & 157 & 0 & -2\, 635 & 55 \\
 4 & -2\, 638 & -201\, 271 & -2\, 638 & 1 & 108\, 874 & 770 \\
 5 & 52\, 029 & 11\, 347\, 905 & 52\, 029 & 0 & -5\, 285\, 162 & 12\, 650 \\
 6 & -1\, 133\, 950 & -708\, 286\, 791 & -1\, 133\, 950 & -2 & 284\, 005\, 066 & 229\, 427 \\
 7 & 26499872 & 47\, 426\, 316\, 644 & 26\, 499\, 872 & 0 & -16\, 372\, 244\, 568 & 4\, 461\, 611 \\
 8 & -651\, 831\, 508 & -3\, 343\, 261\, 431\, 701 & -651\, 831\, 508 & 7 & 993\, 643\, 037\, 662 & 91\, 302\, 244 \\
 \, 9 \, & \, 16\, 673\, 241\, 018 \, & \, 245\, 124\, 281\, 115\, 981 \, & \, 16\, 673\, 241\, 018 \, & \, 0\,  & \, -62\, 721\, 577\, 459\, 427 \, & \, 1\, 942\, 795\, 668 \, \\
\hline  
\end{array}
\ee  
\caption{Extremal BPS invariant for the $10_{124}$, $10_{132}$ and $10_{139}$ knots. 
}      \label{10 br_min_max_from_aug}
\end{table}


\section{LMOV invariants for various knots}   \label{app-lmov}

\begin{table}[H]
\be
\begin{array}{|c|ccccc|}
\hline 
i \setminus j & -4 & -2 & 0 & 2 & 4 \\
\hline 
-5 & 0 & -1 & 1 & -1 & 0 \\
-3 & 1 & 0 & 1 & 0 & 1 \\
 -1 & -1 & 0 & -2 & 0 & -1 \\
 1 & 0 & 1 & 0 & 1 & 0 \\
\hline  
\end{array}
\ee  
\caption{LMOV invariants $N_{1,i,j}$ for the $6_2$ knot. 
}      \label{62-N1ij}
\end{table}

\begin{table}[H]
\be
\begin{array}{|c|cccccccccccc|}
\hline 
i \setminus j & -13 & -11 & -9 & -7 & -5 & -3 & -1 & 1 & 3 & 5 & 7 & 9 \\
\hline 
-10 &  0 & -1 & 1 & -1 & -1 & 1 & -1 & 0 & 0 & 0 & 0 & 0 \\
-8 &  1 & 1 & 0 & 3 & 1 & 1 & 1 & 2 & -1 & 1 & 0 & 0 \\
-6 &  -2 & 0 & -3 & -3 & -2 & -3 & -2 & -2 & 0 & -1 & 0 & 0 \\
-4 &  1 & 1 & 2 & 2 & 4 & 0 & 4 & -1 & 1 & -1 & 0 & -1 \\
-2 &  0 & -1 & 0 & -2 & -1 & 0 & -1 & 1 & 2 & 1 & 1 & 2 \\
0 &  0 & 0 & 0 & 1 & -1 & 1 & -1 & -1 & -2 & 0 & -2 & -1 \\
2 &  0 & 0 & 0 & 0 & 0 & 0 & 0 & 1 & 0 & 0 & 1 & 0 \\
\hline  
\end{array}
\ee  
\caption{LMOV invariants $N_{2,i,j}$ for the $6_2$ knot. 
}      \label{62-N2ij}
\end{table}

\begin{table}[H]
\begin{small}
\be
\begin{array}{|c|ccccccccccccccccccccccccc|}
\hline 
i \setminus j & -30 & -28 & -26 & -24 & -22 & -20 & -18 & -16 & -14 & -12 & -10 & -8 & -6 & -4 & -2 & 0 & 2 & 4 & 6 & 8 & 10 & 12 & 14 & 16 & 18 \\
\hline 
-15 & 0 & -1 & 1 & -1 & -1 & -1 & 0 & -1 & -1 & -1 & -2 & 0 & 0 & -2 & 1 & -1 & 0 & 0 & 0 & 0 & 0 & 0 & 0 & 0 & 0 \\
-13 &  1 & 1 & 1 & 2 & 5 & 4 & 4 & 5 & 7 & 8 & 7 & 3 & 6 & 4 & 3 & 0 & 3 & -1 & 1 & 0 & 0 & 0 & 0 & 0 & 0 \\
-11 &  -2 & -1 & -4 & -4 & -10 & -9 & -13 & -14 & -19 & -21 & -16 & -16 & -16 & -10 & -9 & -3 & -4 & -1 & -1 & 0 & 0 & 0 & 0 & 0 & 0 \\
-9 &  1 & 3 & 2 & 7 & 10 & 14 & 17 & 24 & 27 & 30 & 28 & 28 & 20 & 22 & 6 & 11 & 0 & 4 & -2 & 1 & -1 & 0 & 0 & 0 & 0 \\
-7 &  0 & -2 & -1 & -5 & -6 & -12 & -14 & -22 & -24 & -27 & -29 & -23 & -20 & -15 & -6 & -3 & -1 & 3 & 1 & 0 & 1 & 0 & -1 & 0 & 0 \\
-5 &  0 & 0 & 1 & 1 & 3 & 4 & 9 & 10 & 14 & 15 & 16 & 11 & 11 & -2 & 4 & -7 & -2 & -7 & 2 & 0 & 2 & 2 & 2 & 2 & 0 \\
-3 &  0 & 0 & 0 & 0 & -1 & 0 & -3 & -3 & -4 & -5 & -5 & -3 & 2 & 2 & 5 & 3 & 7 & 0 & -3 & -6 & -5 & -5 & -4 & -4 & -1 \\
-1 &  0 & 0 & 0 & 0 & 0 & 0 & 0 & 1 & 0 & 1 & 1 & 0 & -3 & 0 & -3 & -1 & -1 & 3 & 6 & 8 & 6 & 5 & 6 & 3 & 2 \\
1 &  0 & 0 & 0 & 0 & 0 & 0 & 0 & 0 & 0 & 0 & 0 & 0 & 0 & 1 & -1 & 0 & -2 & -2 & -5 & -4 & -4 & -3 & -3 & -2 & -1 \\
3 &  0 & 0 & 0 & 0 & 0 & 0 & 0 & 0 & 0 & 0 & 0 & 0 & 0 & 0 & 0 & 1 & 0 & 1 & 1 & 1 & 1 & 1 & 0 & 1 & 0 \\
\hline  
\end{array}
\ee  
\end{small}
\caption{LMOV invariants $N_{3,i,j}$ for the $6_2$ knot. 
}      \label{62-N3ij}
\end{table}

\begin{table}[H]
\be
\begin{array}{|c|ccccc|}
\hline 
i \setminus j & -4 & -2 & 0 & 2 & 4 \\
\hline 
-3 &  0 & 1 & -1 & 1 & 0 \\
 -1 &  -1 & 0 & -2 & 0 & -1 \\
 1 &  1 & 0 & 2 & 0 & 1 \\
 3 &  0 & -1 & 1 & -1 & 0 \\
\hline  
\end{array}
\ee  
\caption{LMOV invariants $N_{1,i,j}$ for the $6_3$ knot. 
}      \label{63-N1ij}
\end{table}

\begin{table}[H]
\be
\begin{array}{|c|cccccccccccc|}
\hline 
i \setminus j & -11 & -9 & -7 & -5 & -3 & -1 & 1 & 3 & 5 & 7 & 9 & 11\\
\hline 
-6 & 0 & -1 & 1 & 0 & -1 & 0 & 1 & -1 & 0 & 0 & 0 & 0 \\
-4 &  1 & 1 & -1 & 1 & 3 & -3 & 2 & 0 & 0 & 0 & 0 & 0 \\
-2 &  -2 & 0 & -1 & -3 & -1 & 1 & -2 & 1 & 1 & 0 & 1 & 0 \\
0 &  1 & 1 & 1 & 3 & -1 & 3 & -3 & 1 & -3 & -1 & -1 & -1 \\
2 &  0 & -1 & 0 & -1 & -1 & 2 & -1 & 1 & 3 & 1 & 0 & 2 \\
4 &  0 & 0 & 0 & 0 & 0 & -2 & 3 & -3 & -1 & 1 & -1 & -1 \\
6 &  0 & 0 & 0 & 0 & 1 & -1 & 0 & 1 & 0 & -1 & 1 & 0 \\
\hline  
\end{array}
\ee  
\caption{LMOV invariants $N_{2,i,j}$ for the $6_3$ knot. 
}      \label{63-N2ij}
\end{table}

\begin{table}[H]
\begin{small}
\be
\begin{array}{|c|ccccccccccccccccccccccccc|}
\hline 
i \setminus j & -24 & -22 & -20 & -18 & -16 & -14 & -12 & -10 & -8 & -6 & -4 & -2 & 0 & 2 & 4 & 6 & 8 & 10 & 12 & 14 & 16 & 18 & 20 & 22 & 24 \\
\hline 
-9 & 0 & 1 & -1 & 0 & 0 & 1 & 0 & -1 & 1 & 0 & -1 & 1 & 0 & 0 & 0 & 0 & 0 & 0 & 0 & 0 & 0 & 0 & 0 & 0 & 0 \\
-7 &  -1 & -1 & 0 & 1 & -2 & -3 & -1 & 1 & -1 & -1 & 0 & -3 & 3 & -2 & 0 & 0 & 0 & 0 & 0 & 0 & 0 & 0 & 0 & 0 & 0 \\
-5 &  2 & 1 & 2 & 0 & 6 & 6 & 4 & 3 & 2 & 4 & 7 & -4 & 1 & 1 & 1 & -1 & 0 & 0 & 0 & 0 & 0 & 0 & 0 & 0 & 0 \\
-3 &  -1 & -3 & -1 & -4 & -7 & -10 & -8 & -7 & -8 & -8 & -6 & 5 & -6 & 5 & -4 & 2 & 0 & -2 & -1 & 0 & -1 & 0 & 0 & 0 & 0 \\
-1 &  0 & 2 & 1 & 4 & 5 & 9 & 9 & 8 & 9 & 8 & -2 & 8 & -8 & 3 & 1 & 2 & 3 & 6 & 5 & 3 & 3 & 1 & 1 & 0 & 0 \\
1 &  0 & 0 & -1 & -1 & -3 & -3 & -5 & -6 & -3 & -2 & -1 & -3 & 8 & -8 & 2 & -8 & -9 & -8 & -9 & -9 & -5 & -4 & -1 & -2 & 0 \\
3 &  0 & 0 & 0 & 0 & 1 & 0 & 1 & 2 & 0 & -2 & 4 & -5 & 6 & -5 & 6 & 8 & 8 & 7 & 8 & 10 & 7 & 4 & 1 & 3 & 1 \\
5 &  0 & 0 & 0 & 0 & 0 & 0 & 0 & 0 & 0 & 1 & -1 & -1 & -1 & 4 & -7 & -4 & -2 & -3 & -4 & -6 & -6 & 0 & -2 & -1 & -2 \\
7 &  0 & 0 & 0 & 0 & 0 & 0 & 0 & 0 & 0 & 0 & 0 & 2 & -3 & 3 & 0 & 1 & 1 & -1 & 1 & 3 & 2 & -1 & 0 & 1 & 1 \\
9 &  0 & 0 & 0 & 0 & 0 & 0 & 0 & 0 & 0 & 0 & 0 & 0 & 0 & -1 & 1 & 0 & -1 & 1 & 0 & -1 & 0 & 0 & 1 & -1 & 0 \\
\hline  
\end{array}
\ee  
\end{small}
\caption{LMOV invariants $N_{3,i,j}$ for the $6_3$ knot. 
}      \label{63-N3ij}
\end{table}

\begin{table}[H]
\be
\begin{array}{|c|ccccc|}
\hline 
i \setminus j & -4 & -2 & 0 & 2 & 4 \\
\hline 
3 &  -1 & 1 & -1 & 1 & -1 \\
5 &  0 & 0 & -1 & 0 & 0 \\
7 &  1 & 0 & 2 & 0 & 1 \\
9 &  0 & -1 & 0 & -1 & 0 \\
\hline  
\end{array}
\ee  
\caption{LMOV invariants $N_{1,i,j}$ for the $7_3$ knot. 
}      \label{73-N1ij}
\end{table}

\begin{table}[H]
\be
\begin{array}{|c|ccccccccccccc|}
\hline 
i \setminus j & -5 & -3 & -1 & 1 & 3 & 5 & 7 & 9 & 11 & 13 & 15 & 17 & 19\\
\hline 
6 &  0 & 1 & -2    & 2 & 0 & 0 & 0 & 1 & -1 & 1 & 0 & 0 & 0 \\
8 &  -1 & 1 & 0    & -1 & -1 & 1 & -1 & 0 & 1 & -1 & 0 & 0 & 0 \\
10 &  1 & -2 & 2  & 0 & 1 & 0 & 1 & -1 & 0 & 0 & 0 & 0 & 0 \\
12 &  0 & -2 & 0  & -5 & -3 & -4 & -4 & -3 & -2 & -2 & -2 & 0 & -1 \\
14 &  1 & 2 & 4    & 6 & 8 & 7 & 10 & 6 & 6 & 5 & 4 & 1 & 2 \\
16 &  -1 & -1 & -4 & -4 & -6 & -7 & -7 & -5 & -5 & -5 & -2 & -2 & -1 \\
18 &  0 & 1 & 0    & 2 & 1 & 3 & 1 & 2 & 1 & 2 & 0 & 1 & 0 \\
\hline  
\end{array}
\ee  
\caption{LMOV invariants $N_{2,i,j}$ for the $7_3$ knot. 
}      \label{73-N2ij}
\end{table}

\begin{table}[H]
\be
\begin{array}{|c|ccccc|}
\hline 
i \setminus j & -4 & -2 & 0 & 2 & 4 \\
\hline 
-9 &  0 & 1 & -1 & 1 & 0 \\
-7 & -1 & 1 & -1 & 1 & -1 \\
-5 &  0 & -1 & 0 & -1 & 0 \\
-3 &  1 & -1 & 2 & -1 & 1 \\
\hline  
\end{array}
\ee  
\caption{LMOV invariants $N_{1,i,j}$ for the $7_5$ knot. 
}      \label{75-N1ij}
\end{table}

\begin{table}[H]
\be
\begin{array}{|c|ccccccccccccc|}
\hline 
i \setminus j & -19 & -17 & -15 & -13 & -11 & -9 & -7 & -5 & -3 & -1 & 1 & 3 & 5\\
\hline 
-18 &  0 & -1 & 1 & -1 & -1 & 0 & 0 & -1 & 0 & -1 & 1 & -1 & 0 \\
-16 &  1 & 1 & -1 & 3 & 2 & -1 & 2 & 1 & 1 & -1 & 2 & -1 & 1 \\
-14 &  -2 & 1 & -1 & -3 & 0 & 0 & -2 & 0 & 1 & -1 & 1 & 0 & 0 \\
-12 &  1 & -1 & 1 & 1 & 0 & 1 & 1 & 1 & -2 & 4 & -4 & 2 & -1 \\
-10 &  0 & 0 & 0 & -1 & -2 & 0 & -5 & -1 & -3 & 0 & -5 & 3 & -2 \\
-8 &  0 & 0 & 0 & 2 & 0 & 2 & 5 & 0 & 3 & 3 & 2 & -1 & 2 \\
-6 &  0 & 0 & 0 & -1 & 1 & -2 & -1 & 0 & 0 & -4 & 3 & -2 & 0 \\
\hline  
\end{array}
\ee  
\caption{LMOV invariants $N_{2,i,j}$ for the $7_5$ knot. 
}      \label{75-N2ij}
\end{table}

\begin{table}[H]
\be
\begin{array}{|c|ccccccccc|}
\hline 
i \setminus j & -8 & -6 & -4 & -2 & 0 & 2 & 4 & 6 & 8 \\
\hline 
-13 &  0 & 0 & 0 & -1 & 0 & -1 & 0 & 0 & 0 \\
-11 &  0 & 1 & 1 & 2 & 2 & 2 & 1 & 1 & 0 \\
-9 & -1 & -1 & -2 & -2 & -3 & -2 & -2 & -1 & -1 \\
-7 &  1 & 0 & 1 & 1 & 1 & 1 & 1 & 0 & 1 \\
\hline  
\end{array}
\ee  
\caption{LMOV invariants $N_{1,i,j}$ for the $10_{124}$ knot. 
}      \label{10124-N1ij}
\end{table}

\begin{table}[H]
\begin{tiny}
\be
\begin{array}{|c|ccccccccccccccccccccccc|}
\hline 
i \setminus j & -31 & -29 & -27 & -25 & -23 & -21 & -19 & -17 & -15 & -13 & -11 & -9 & -7 & -5 & -3 & -1 & 1 & 3 & 5 & 7 & 9 & 11 & 13\\
\hline 
-26 &  0 & 0 & 0 & 0 & -1 & 0 & -2 & -1 & -2 & -1 & -3 & -1 & -3 & -1 & -2 & -1 & -1 & 0 & 0 & 0 & 0 & 0 & 0 \\
-24 &  0 & 0 & 1 & 2 & 4 & 6 & 9 & 11 & 13 & 15 & 17 & 18 & 18 & 17 & 15 & 12 & 8 & 5 & 2 & 1 & 0 & 0 & 0 \\
-22 &  0 & -2 & -3 & -9 & -12 & -22 & -26 & -39 & -41 & -55 & -55 & -65 & -60 & -63 & -51 & -46 & -31 & -23 & -12 & -7 & -2 & -1 & 0 \\
-20 &  1 & 4 & 7 & 15 & 24 & 35 & 49 & 64 & 77 & 92 & 103 & 109 & 112 & 107 & 96 & 81 & 63 & 44 & 29 & 16 & 8 & 3 & 1 \\
-18 &  -2 & -3 & -9 & -13 & -25 & -32 & -50 & -58 & -79 & -85 & -103 & -102 & -111 & -100 & -97 & -77 & -66 & -45 & -33 & -18 & -11 & -4 & -2 \\
-16 &  1 & 2 & 4 & 7 & 12 & 17 & 24 & 31 & 38 & 45 & 50 & 53 & 54 & 52 & 47 & 41 & 33 & 25 & 17 & 11 & 6 & 3 & 1 \\
-14 &  0 & -1 & 0 & -2 & -2 & -4 & -4 & -8 & -6 & -11 & -9 & -12 & -10 & -12 & -8 & -10 & -6 & -6 & -3 & -3 & -1 & -1 & 0 \\
\hline  
\end{array}
\ee  
\end{tiny}
\caption{LMOV invariants $N_{2,i,j}$ for the $10_{124}$ knot. 
}      \label{10124-N2ij}
\end{table}

\begin{table}[H]
\be
\begin{array}{|c|ccccc|}
\hline 
i \setminus j & -4 & -2 & 0 & 2 & 4 \\
\hline 
-7 &  0 & 1 & 0 & 1 & 0 \\
-5 &  -1 & -1 & -1 & -1 & -1 \\
-3 &  1 & 0 & 1 & 0 & 1 \\
\hline  
\end{array}
\ee  
\caption{LMOV invariants $N_{1,i,j}$ for the $10_{132}$ knot. 
}      \label{10132-N1ij}
\end{table}

\begin{table}[H]
\be
\begin{array}{|c|cccccccccccccc|}
\hline 
i \setminus j & -19 & -17 & -15 & -13 & -11 & -9 & -7 & -5 & -3 & -1 & 1 & 3 & 5 & 7\\
\hline 
-14 &  0 & -1 & 0 & -1 & -1 & -2 & -1 & -3 & -1 & -2 & 0 & -1 & 0 & 0 \\
-12 &  1 & 2 & 2 & 4 & 5 & 7 & 8 & 8 & 7 & 6 & 4 & 2 & 1 & 0 \\
-10 &  -2 & -2 & -5 & -6 & -10 & -11 & -15 & -11 & -14 & -9 & -8 & -3 & -3 & 0 \\
-8 &  1 & 2 & 4 & 5 & 9 & 10 & 11 & 11 & 11 & 8 & 6 & 4 & 2 & 1 \\
-6 &  0 & -1 & -1 & -3 & -3 & -5 & -4 & -6 & -3 & -4 & -2 & -2 & 0 & -1 \\
-4 &  0 & 0 & 0 & 1 & 0 & 1 & 1 & 1 & 0 & 1 & -1 & 0 & 0 & -1 \\
-2 &  0 & 0 & 0 & 0 & 0 & 0 & 0 & 0 & 0 & 1 & 0 & 0 & 1 & 1 \\
0 &  0 & 0 & 0 & 0 & 0 & 0 & 0 & 0 & 0 & -1 & 1 & 0 & -1 & 0 \\
\hline  
\end{array}
\ee  
\caption{LMOV invariants $N_{2,i,j}$ for the $10_{132}$ knot. 
}      \label{10132-N2ij}
\end{table}

\begin{table}[H]
\be
\begin{array}{|c|ccccccccc|}
\hline 
i \setminus j & -8 & -6 & -4 & -2 & 0 & 2 & 4 & 6 & 8 \\
\hline 
7 &  -1 & 0 & -1 & -1 & 0 & -1 & -1 & 0 & -1 \\
9 &  1 & 1 & 2 & 1 & 2 & 1 & 2 & 1 & 1 \\
11 &  0 & -1 & -1 & -1 & -1 & -1 & -1 & -1 & 0 \\
13 &  0 & 0 & 0 & 1 & -1 & 1 & 0 & 0 & 0 \\
\hline  
\end{array}
\ee  
\caption{LMOV invariants $N_{1,i,j}$ for the $10_{139}$ knot. 
}      \label{10139-N1ij}
\end{table}

\begin{table}[H]
\begin{small}
\be
\begin{array}{|c|ccccccccccccccccccccccc|}
\hline 
i \setminus j & -13 & -11 & -9 & -7 & -5 & -3 & -1 & 1 & 3 & 5 & 7 & 9 & 11 & 13 & 15 & 17 & 19 & 21 & 23 & 25 & 27 & 29 & 31\\
\hline 
14 &  0 & 1 & 1 & 2 & 2 & 4 & 4 & 7 & 5 & 8 & 8 & 8 & 6 & 9 & 4 & 7 & 3 & 3 & 2 & 2 & 0 & 1 & 0 \\
16 &  -1 & -3 & -5 & -8 & -11 & -16 & -21 & -25 & -29 & -33 & -35 & -34 & -33 & -32 & -27 & -23 & -18 & -13 & -10 & -6 & -4 & -2 & -1 \\
18 &  2 & 4 & 9 & 13 & 21 & 27 & 38 & 43 & 54 & 57 & 63 & 59 & 63 & 53 & 52 & 38 & 35 & 22 & 19 & 10 & 8 & 3 & 2 \\
20 &  -1 & -3 & -7 & -11 & -18 & -24 & -32 & -41 & -46 & -52 & -56 & -54 & -55 & -51 & -43 & -38 & -30 & -21 & -16 & -11 & -5 & -4 & -1 \\
22 &  0 & 1 & 2 & 5 & 7 & 11 & 14 & 19 & 21 & 25 & 24 & 27 & 24 & 26 & 18 & 20 & 13 & 12 & 6 & 6 & 2 & 2 & 0 \\
24 &  0 & 0 & 0 & -1 & -1 & -2 & -3 & -4 & -5 & -5 & -5 & -6 & -6 & -5 & -4 & -5 & -4 & -2 & -2 & -1 & -1 & 0 & 0 \\
26 &  0 & 0 & 0 & 0 & 0 & 0 & 0 & 1 & 0 & 0 & 1 & 0 & 1 & 0 & 0 & 1 & 1 & -1 & 1 & 0 & 0 & 0 & 0 \\
\hline  
\end{array}
\ee  
\end{small}
\caption{LMOV invariants $N_{2,i,j}$ for the $10_{139}$ knot. 
}      \label{10139-N2ij}
\end{table}



\newpage

\bibliographystyle{halpha}
\bibliography{abmodel}

\end{document}